\documentclass[]{emulateapj}

\usepackage{graphics,graphicx,xspace,natbib,amssymb}
\usepackage[caption=false]{subfig}
\usepackage{amsmath}
\usepackage{threeparttable}

\newcommand{\sersic}{S\'{e}rsic\xspace}

\shorttitle{Galaxy half-mass radii}
\shortauthors{Suess et al.}

\begin{document}

\title{Half-mass radii for $\sim7,000$ galaxies at $1.0\le \lowercase{z}\le2.5$:
most of the evolution in the mass-size relation is due to color gradients}
\author{Katherine A. Suess\altaffilmark{1}, Mariska Kriek\altaffilmark{1}, Sedona H. Price\altaffilmark{2}, Guillermo Barro\altaffilmark{3}} 

\altaffiltext{1}{Astronomy Department, University of California, Berkeley, CA 94720, USA}
\altaffiltext{2}{9Max-Planck-Institut f{\"u}r extraterrestrische Physik, Postfach 1312, Garching, 85741, Germany}
\altaffiltext{3}{Department of Physics, University of the Pacific, 3601 Pacific Ave, Stockton, CA 95211, USA}
\email{suess@berkeley.edu}

\begin{abstract}
Radial mass-to-light ratio gradients cause the half-mass and half-light radii of galaxies to differ, potentially biasing studies that use half-light radii. Here we present the largest catalog to date of galaxy half-mass radii at $z>1$: 7,006 galaxies in the CANDELS fields at $1.0\le z \le 2.5$. The sample includes both star-forming and quiescent galaxies with stellar masses $9.0 \le \log{(\rm{M_*}/\rm{M}_\odot)} \le 11.5$.  We test three methods for calculating half-mass radii from multi-band PSF-matched {\it HST} imaging: two based on spatially-resolved SED modeling, and one that uses a rest-frame color profile. All three methods agree, with scatter $\lesssim 0.3$ dex. In agreement with previous studies, most galaxies in our sample have negative color gradients (the centers are redder than the outskirts, and $r_{\rm{e,mass}} < r_{\rm{e,light}}$). We find that color gradient strength has significant trends with increasing stellar mass, half-light radius, $U-V$ color, and stellar mass surface density. These trends have not been seen before at $z>1$. Furthermore, color gradients of star-forming and quiescent galaxies show a similar redshift evolution: they are flat at $z\gtrsim2$, then steeply decrease as redshift decreases. This affects the galaxy mass-size relation. The normalizations of the star-forming and quiescent $r_{\rm{mass}}-M_*$ relations are 10-40\% smaller than the corresponding $r_{\rm{light}}-M_*$ relations; the slopes are $\sim0.1-0.3$ dex shallower. Finally, the half-mass radii of star-forming and quiescent galaxies at $M_* = 10^{10.5}M_\odot$  only grow by $\sim 1\%$ and $\sim8\%$ between $z\sim2.25$ and $z\sim1.25$. This is significantly less than the $\sim37\%$ and $\sim47\%$ size increases found when using the half-light radius.
\end{abstract}


%------------------------------------------------------------------------------------------------------------------------------------------------------------------------
\section{Introduction}
The half-light radii of both quiescent and star-forming galaxies are related to their stellar masses, such that massive galaxies are larger than low-mass galaxies.
This galaxy mass-size relation has now been studied over a large range of redshifts and stellar masses \citep[e.g.,][]{shen03,vanderwel14,mowla18}. The mass-size relation for star-forming galaxies has a relatively shallow slope. The sizes of disky star-forming galaxies seem to be proportional to the virial radii of their host dark matter halos, likely a result of conservation of angular momentum as disks form  \citep[e.g.,][]{kravtsov13,huang17,somerville17}.

Quiescent galaxies, on the other hand, exhibit a much steeper slope in the $r_{\rm{e, light}}-M_*$ relation and are more compact than similar-mass star-forming galaxies at all epochs \citep[e.g.,][]{vanderwel14}. Furthermore, it now seems well-established that quiescent galaxies have experienced remarkable size growth over cosmic time, more than doubling their half-light radii between $z\sim2$ and $z\sim0$ \citep[e.g.,][]{daddi05,vandokkum08,damjanov09,szomoru10,damjanov11, vanderwel14}. This extreme size evolution has led to multiple competing interpretations. In the `inside-out growth' scenario, quiescent galaxies grow at late times via minor mergers that increase their radii without significantly increasing their stellar mass \citep[e.g.,][]{bezanson09,naab09,vandesande13,hopkins09}. In the `progenitor bias' scenario, galaxies that quench later are larger, introducing bias into average size evolution \citep[e.g.,][]{vandokkum01,carollo13,poggianti13}. 

However, these studies all rely on galaxy half-light radii. Because light is a biased tracer of mass, half-light radii are {\it not a direct probe} of the underlying stellar mass profiles of galaxies. Differences in half-mass and half-light radii arise from radial gradients in mass-to-light ratios. These radial mass-to-light ratio gradients can be physically caused by radial gradients in stellar populations: older, more metal-poor, or dustier stellar populations have higher mass-to-light ratios than younger or more metal-rich ones. Stellar mass-to-light ratio gradients can be observed as {\it color} gradients-- stellar populations with high mass-to-light ratios are redder than those with low mass-to-light ratios. These color gradients bias studies of galaxy evolution that use half-light radii instead of half-mass radii. 
A robust catalog of galaxy half-mass radii is required to understand the effects that this bias has on studies of galaxy size evolution.

Studying galaxy color gradients also allows us to directly probe the assembly histories of galaxies, because different assembly histories naturally lead to different mass-to-light ratio profiles. For example, inside-out growth results in negative color gradients, as the accreted younger and/or lower-metallicity stellar populations primarily reside in the outskirts of the galaxy \citep[e.g.,][]{naab09}. Similarly, inside-out growth via star formation--- where star formation occurs in the outer disk of a galaxy, but not its central bulge \citep[e.g.,][]{nelson16}--- would result in negative color gradients. A central starburst (without extreme dust obscuration) will result in positive color gradients, where younger stars are found preferentially in the center of the galaxy. Uniform growth at all radii would result in no radial color gradient.
Studying galaxy mass-to-light ratio profiles thus serves two important purposes: first, it gives us a less biased tracer of galaxy mass profiles; second, it probes the assembly histories of galaxies. 

Some work has been done to understand galaxy mass-to-light ratio gradients, both in the $z\sim 0$ and $z>1$ universe.
In the local universe, radial color gradients attributed to metallicity gradients have been observed in quiescent galaxies \citep[e.g.,][]{saglia00, labarbera05,tortora10}. At higher redshifts, where galaxy half-light radii are smaller and observations are more difficult, it is challenging to break the age-metallicity degeneracy; still, negative color gradients implying radial variations in age, metallicity, or dust have been observed in many quiescent galaxies \citep[e.g.,][]{wuyts10, guo11, szomoru12, chan16, liu2017, mosleh17}. 

Previous studies at $z \gtrsim 1$ have typically used one of two methods to study $M/L$ variations within galaxies. The first technique exploits the empirical correlation between rest-frame color and mass-to-light ratio \citep[e.g.,][]{bell01}. A intrinsic rest-frame color profile is constructed for the galaxy using \texttt{GALFIT} \citep{peng02}, converted to a mass-to-light ratio profile, and used to calculate a mass profile and a half-mass radius \citep[e.g.][]{zibetti09, szomoru10,szomoru12,szomoru13,chan16}. The second technique uses multi-band imaging to measure the spatially-resolved spectral energy distribution (SED) of the galaxy; stellar population synthesis modeling of each spatially-resolved SED can recover the mass, age, and metallicity of the region \citep[e.g.,][]{wuyts12,lang14}. After accounting for the effects of the point spread function (PSF), half-mass radii can then be extracted from the mass maps of the galaxy. 

These techniques have different strengths and potential pitfalls, but there has not yet been a direct comparison between these two major methods for calculating half-mass radii. Furthermore, while studies at $z\sim0$ have large sample sizes \citep[typically via the SDSS, e.g.][]{tortora10}, most studies of color gradients in more distant galaxies have focused on relatively small samples ($<200$) of high-mass quiescent galaxies. Only a single study, \citet{mosleh17}, has included a larger sample over a wider stellar mass range. Here, we wish to perform a large, systematic study of the color gradients and half-mass radii of both quiescent and star-forming galaxies over a large range of stellar masses and redshifts. This type of study requires a large, deep, multi-band imaging survey--- like CANDELS \citep{grogin11}--- that has excellent PSF-matching \citep[now available from the 3D-HST collaboration, ][]{skelton14}.

In this paper, we use this remarkable public dataset, in combination with the multi-wavelength medium- and broad-band photometric catalog from ZFOURGE, to calculate the half-mass radii of 7,006 galaxies at $1.0 \le z \le 2.5$ in a stellar mass range of $9.0 \le \log{(\rm{M_*}/\rm{M}_\odot)} \lesssim 11.5$. We compare three different methods for recovering half-mass radii--- two based on resolved SED modeling, and one based on deconvolved rest-frame color gradients. We present trends between the strength of observed mass-to-light ratio gradients in galaxies and other galaxy properties such as stellar mass and \sersic index. Finally, we focus on the implications that these half-mass radii have for the galaxy mass-size relation and its evolution over cosmic time. Throughout this paper, we assume a cosmology of $\Omega_m = 0.3$, $\Omega_\Lambda = 0.7$, and $h=0.7$; all radii are (non-circularized) measurements of the major axis. 

%------------------------------------------------------------------------------------------------------------------------------------------------------------------------
\section{Data and Sample Selection}

\subsection{Sample Selection}

\begin{figure}
    \centering
    \includegraphics[width=.48\textwidth]{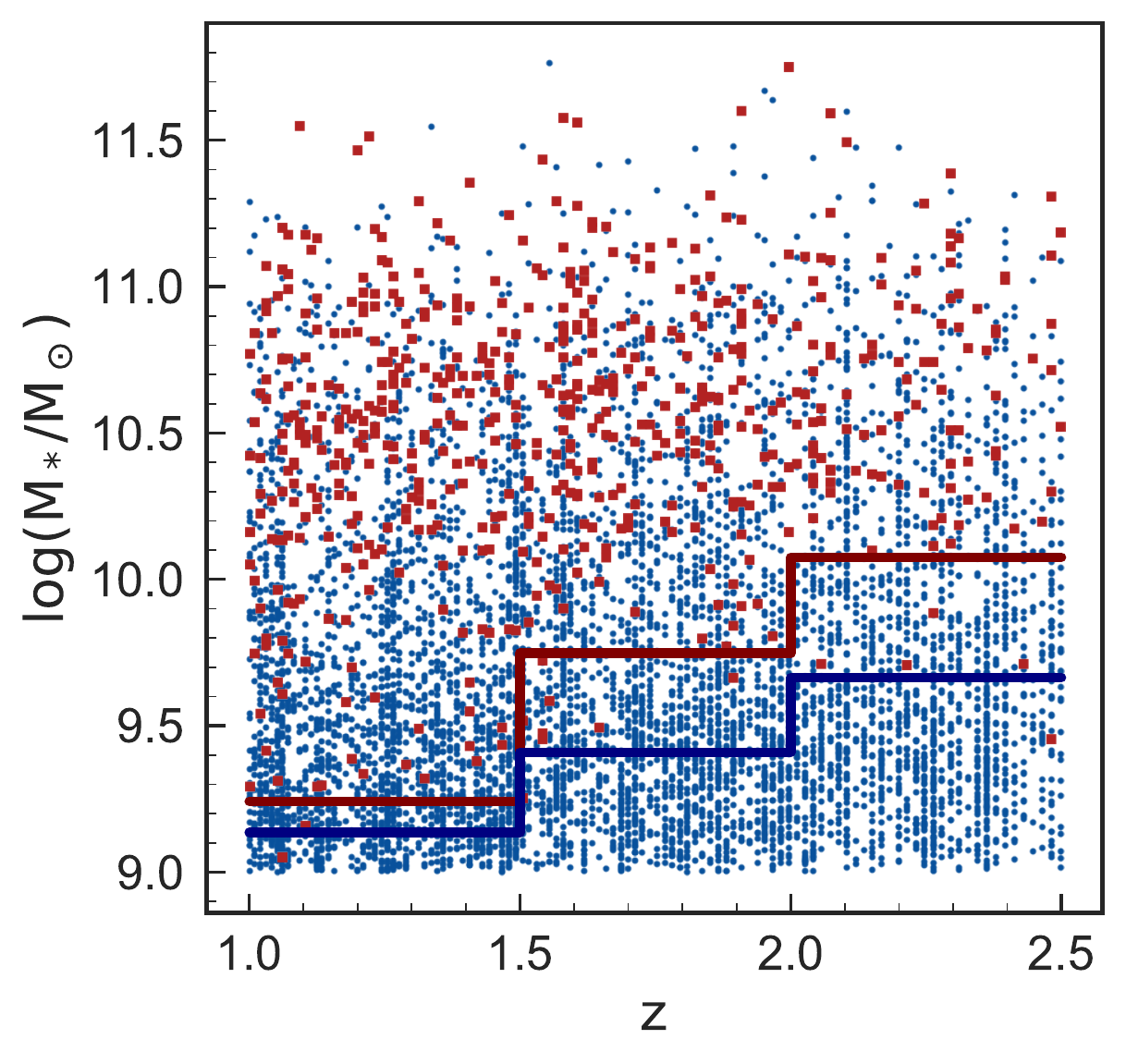}
    \caption{The 7,006 galaxies included in this analysis. Galaxies are classified as star-forming or quiescent by their UVJ colors; quiescent galaxies are shown as red squares, and star-forming galaxies are shown with blue points. The mass completeness in each redshift bin is shown as a red (blue) solid line for quiescent (star-forming) galaxies.}
    \label{fig:sample}
\end{figure}

For our sample selection, we use the multi-wavelength photometric catalogs from ZFOURGE \citep{straatman16}, which overlap with the CANDELS survey in the COSMOS, GOODS-S, and UDS fields. These catalogs contain medium- and broad-band photometry spanning from the optical to IRAC, for a total of 26-40 photometric bands.
Our sample consists of all galaxies in the ZFOURGE catalog with a use flag\footnote{See Table 5 of \citet{straatman16} for a full definition of this flag. The \texttt{use} flag is intended to provide a ``standard selection of galaxies" and ensures that the object is not a star or too close to a star for reliable observations; the object was observed in all optical bands; the object has reasonable EAZY and FAST fits; and that ground- and space-based observations in similar filters have similar fluxes.} equal to one, a reliable photometric or spectroscopic redshift ($\texttt{z\_q < 3}$), $1.0 \le z \le 2.5$, high signal-to-noise ratio in the detection band (S/N$_{\rm{K}}\ge 10$), $\log{M_* / M_{\odot}} > 9.0$, and a match in the 3D-HST catalog \citep[][catalog cross-matching provided as part of the ZFOURGE catalogs]{brammer12,momcheva16,skelton14}. We select from ZFOURGE because the medium bands included as part of the survey provide extra SED coverage, which allows for better stellar population fitting. Because the footprints of the 3D-HST and CANDELS catalogs are slightly different, we also ensure that each galaxy has CANDELS multi-band imaging \citep{grogin11,koekemoer11}. 

We classify each of these 7,006 galaxies as either star-forming or quiescent by their rest-frame $U-V$ vs $V-J$ colors \citep{wuyts07} using the classification from \citet{whitaker12}. A total of 582 galaxies ($\sim8\%$ of the sample) are quiescent. Throughout this work, we use the stellar masses and redshifts as presented in the ZFOURGE catalog. We correct the stellar masses to be consistent with the \citet{vanderwel12} morphological measurements by multiplying the ZFOURGE stellar masses by the ratio of the F160W flux in the \citet{vanderwel12} \texttt{GALFIT} catalog and the F160W flux in the ZFOURGE catalog. On average, this small correction increases the stellar mass by 5\%.

We calculate the mass completeness of both the quiescent and star-forming galaxies using a method similar to those of \citet{quadri12} and \citet{tomczak14}: we select galaxies with a S/N$_{\rm{K}}$ close to our cutoff ($10 \le \rm{S/N}_{\rm{K}} \le 50$), scale their fluxes and masses down to a S/N~$=10$, and take the 90th percentile of the resulting masses as the mass completeness. We calculate the mass completeness in three redshift bins, $1.0 \le z < 1.5$, $1.5 \le z < 2.0$, and $2.0 \le z \le 2.5$. The 90\% completeness level is, in effect, calculated at the center of each redshift bin; the lower-redshift portions of each bin are $>90\%$ complete, and the higher-redshift portions of each bin are $<90\%$ complete.

Figure \ref{fig:sample} shows our full sample of galaxies in stellar mass-redshift space. Star-forming and quiescent galaxies are indicated with blue points and red squares, and the mass completeness in each redshift bin is shown with a solid line. For both quiescent and star-forming galaxies at all redshifts, the sample is complete above $\sim10^{10}M_\odot$.

\subsection{HST Imaging \& Morphologies}
We make extensive use of the high-resolution multi-band imaging and integrated photometry in the COSMOS, UDS, and GOODS-S fields obtained by the CANDELS program \citep{grogin11,koekemoer11}. For this analysis, we use the PSF-matched images created by the 3D-HST team \citep{skelton14,momcheva16}. 

We adopt the half-light radii and \sersic indices for each galaxy from the morphological catalogs presented in \citet{vanderwel12}. These half-light radii, like the half-mass radii presented later in this work, are measurements of the galaxy's major axis. We correct all half-light radii to rest-frame 5,000\AA\ using the procedure described in \citet{vanderwel14}. This technique applies a mass- and redshift-dependent correction to the measured half-light radii of star-forming galaxies. A redshift-dependent correction is applied in a similar fashion to quiescent galaxies.

%------------------------------------------------------------------------------------------------------------------------------------------------------------------------
\section{Methods}
\label{sec:methods}
\begin{figure*}
    \centering
    \includegraphics[width=.9\textwidth]{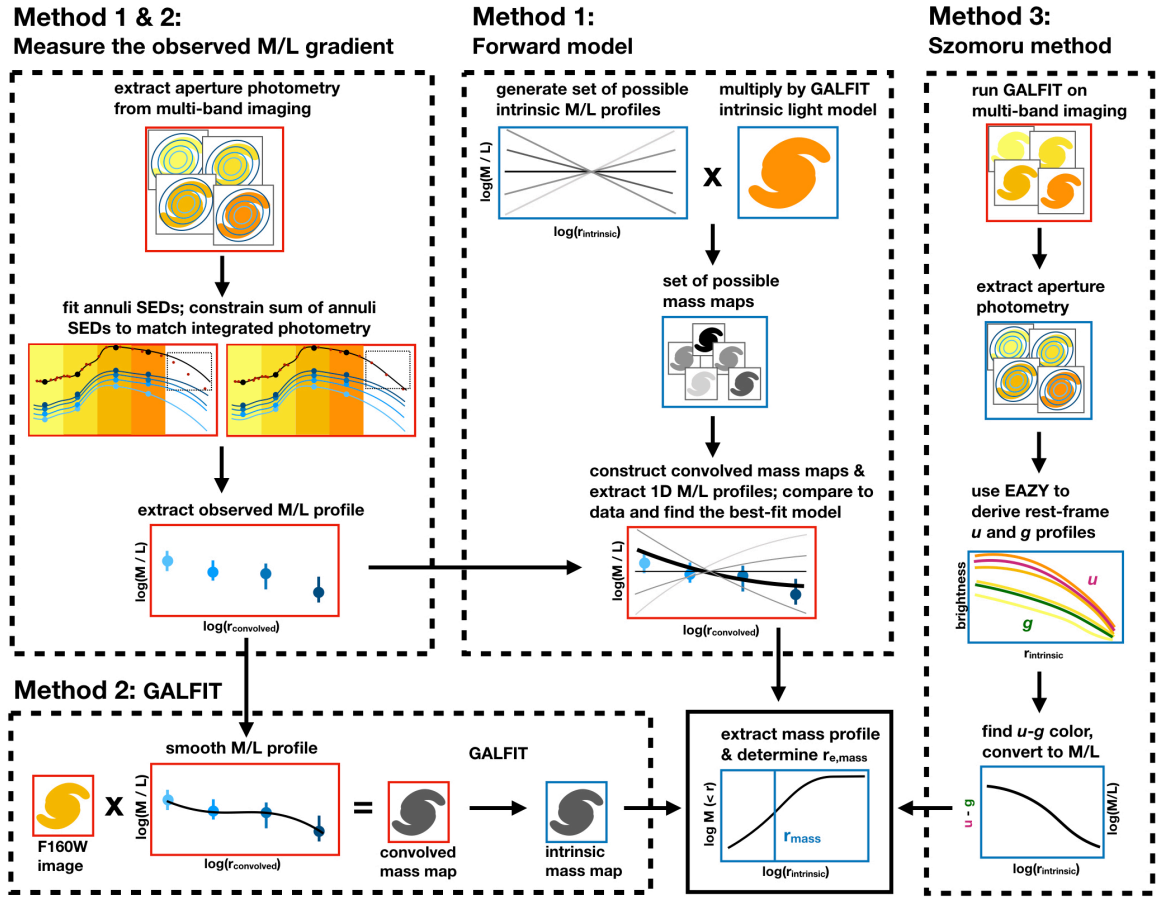}
    \caption{Graphic representation of the three methods used to calculate half-mass radii. Individual steps enclosed with a red box indicate that the measurements are in {\it convolved} space, and steps enclosed with a blue box indicate that the measurements are in deconvolved space.}
    \label{fig:process}
\end{figure*}

\begin{figure*}
    \centering
    \subfloat{{\includegraphics[width=.32\textwidth]{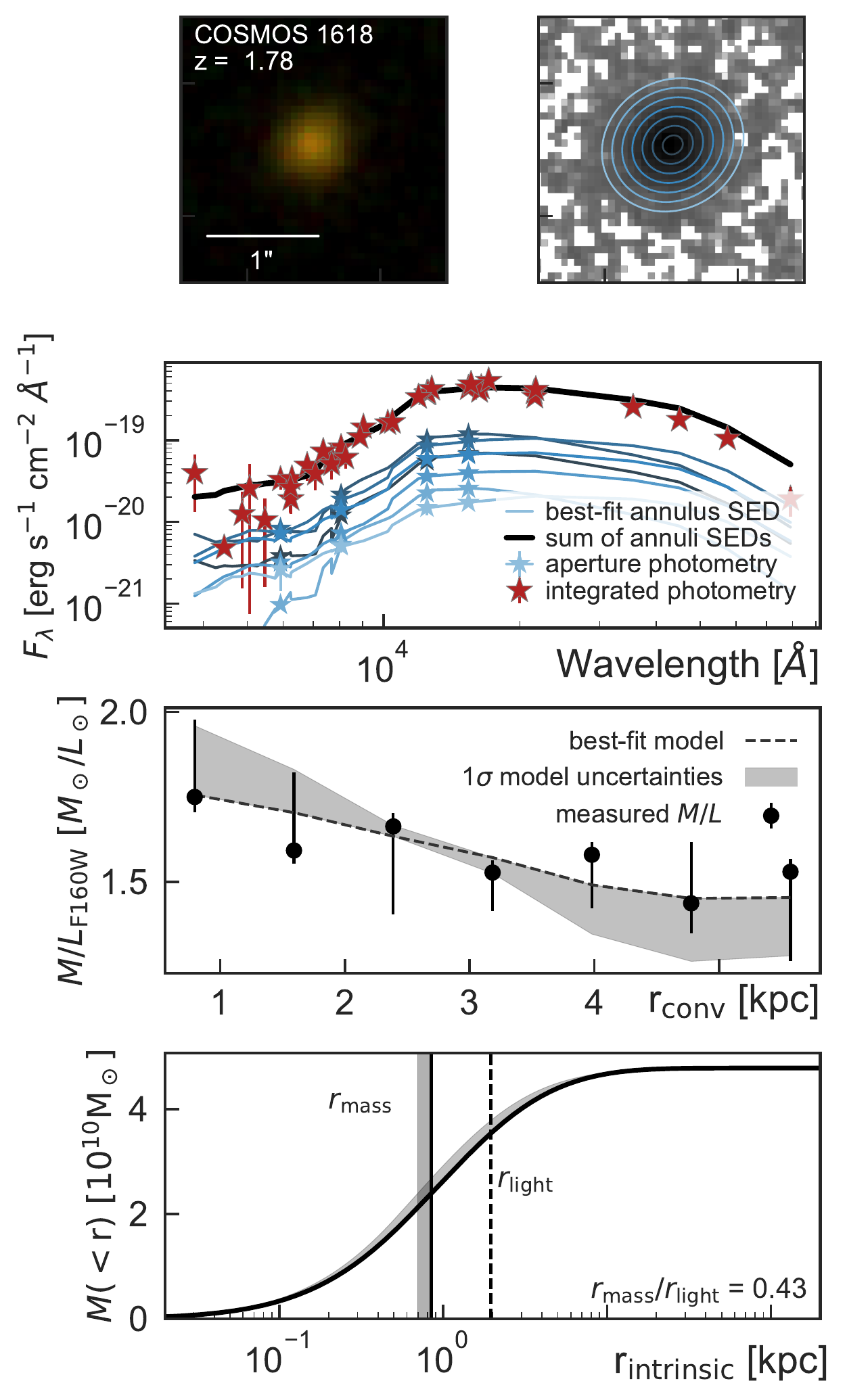} }}
    \subfloat{{\includegraphics[width=.32\textwidth]{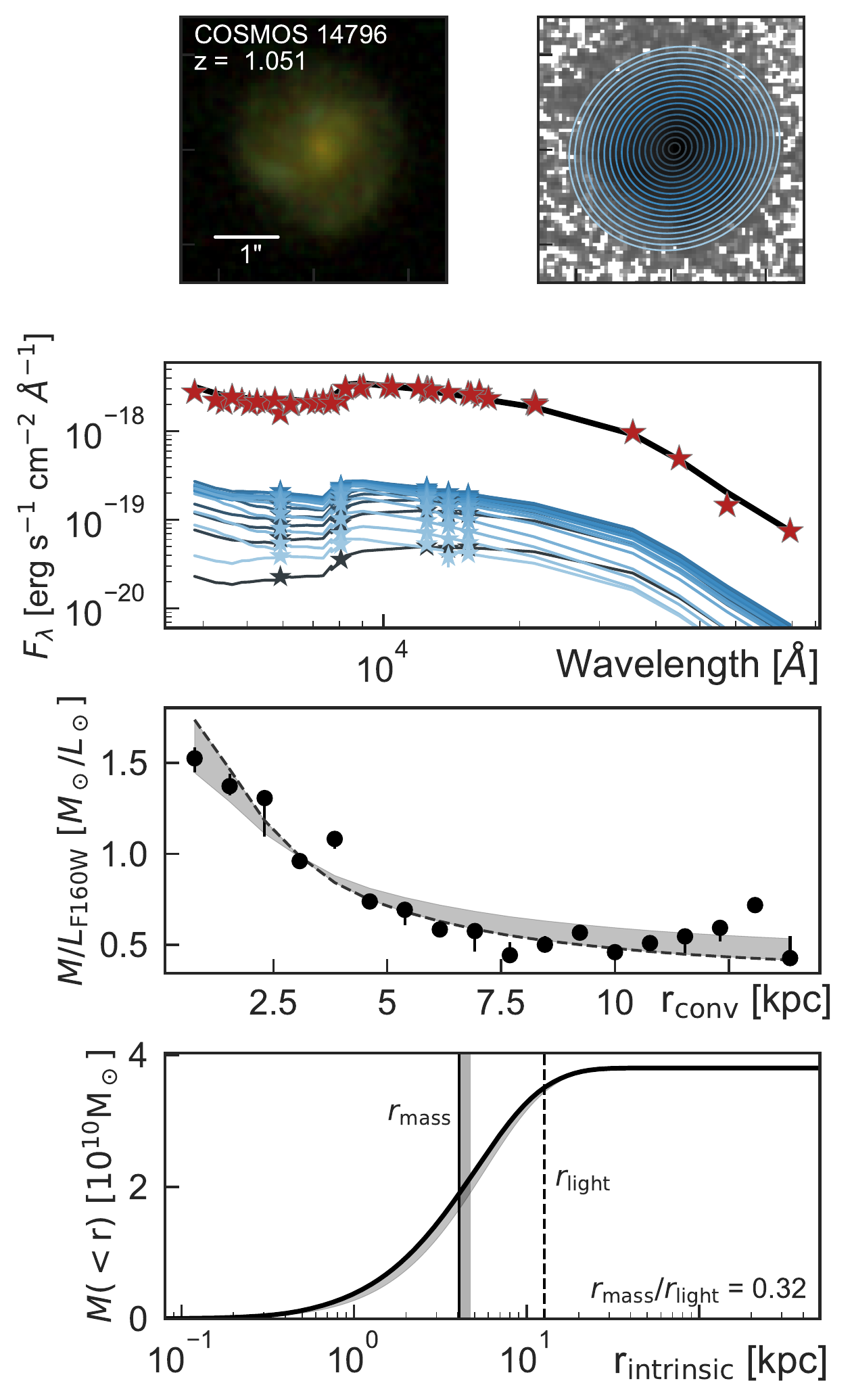} }}
    \subfloat{{\includegraphics[width=.32\textwidth]{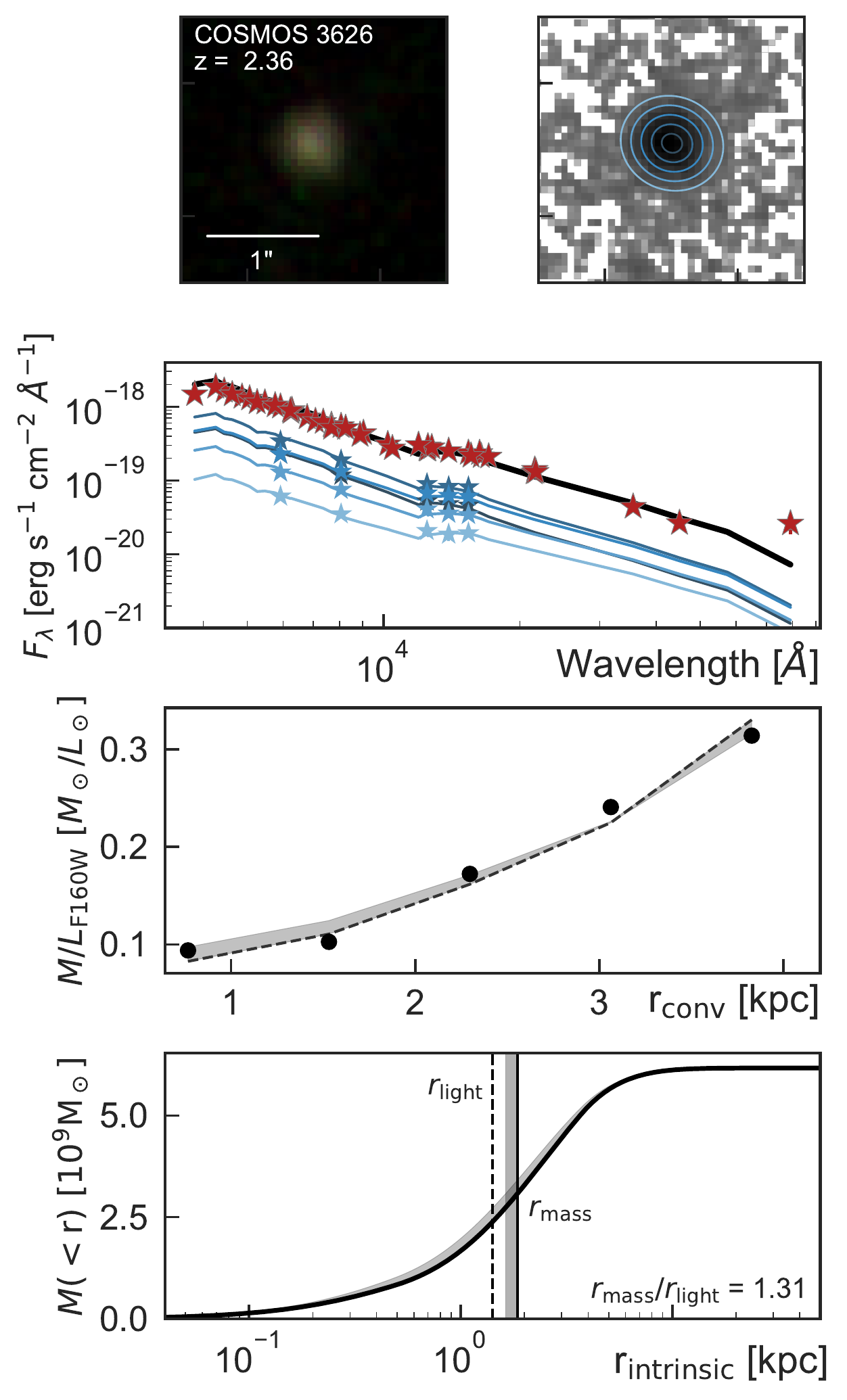} }}
    \caption{Examples of Method 1 for three galaxies in our sample. The top row shows a color image of the galaxy, as well as a log-scaled F160W image with the locations of the annuli used to extract aperture photometry. The second row shows the measured aperture photometry in each band (blue stars), the best-fit SED for each aperture (blue lines), the sum of all best-fit annuli SEDs (black line), and the integrated catalog photometry for the whole galaxy (red stars). The shade of blue used for the aperture photometry and the best-fit annulus SED matches the shade used in the top panel image (dark blue are inner annuli, light blue are outer annuli). The third row shows the measured $M/L$ profile as a function of radius (in convolved space). Points represent the measured values for each annulus; the black dashed line shows the best-fit power-law model, and the grey shaded region shows the $1\sigma$ error bars on the best-fit model. The bottom row shows the mass profile corresponding to this best-fit model, as a function of {\it intrinsic} radius. The vertical line and shaded region represent the half-mass radius and its 1 $\sigma$ error bar; the dashed line shows the half-light radius.}
    \label{fig:method1}
\end{figure*}

In this paper, we explore three different methods to calculate half-mass radii. We introduce a new technique that builds on the methods used by \citet{wuyts12}: we divide the galaxy into annuli, measure multi-band aperture photometry in each annulus, then use stellar population synthesis (SPS) modeling to fit the resulting SEDs and obtain a mass map of the galaxy. We then use two separate techniques to account for the {\it HST} PSF and derive the intrinsic mass profiles. The first approach uses a forward modeling technique that assumes the mass-to-light ratio gradient is a power-law function of radius; the second approach uses \texttt{GALFIT} \citep{peng02} to fit the mass map of the galaxy \citep[similar to the technique used by][]{lang14, chan16}. Our third and final method replicates the analysis of \citet{szomoru10,szomoru12,szomoru13}, and uses the rest-frame intrinsic $u-g$ color profile to create a mass profile and measure a half-mass radius. The following sections describe each of these methods in greater detail. Figure \ref{fig:process} shows a graphical representation of the major steps for each method, and Figure \ref{fig:method1} shows example images and mass profiles for three galaxies in our sample. Similar to \citet{vanderwel14} and \citet{mowla18}, all half-mass and half-light radii presented in this paper are measurements of the major axis of the galaxy.

\subsection{Methods 1 \& 2: extracting the $M/L$ gradient in convolved space}
\label{mainmethod}
We begin by dividing each galaxy into distinct spatial regions. Because we aim to measure the radial mass profiles of galaxies, we define these spatial regions as concentric elliptical annuli; this geometry allows us to extract radial profiles much more simply than if we define regions using the Voronoi binning technique used by \citet{wuyts12}. The annuli are centered at the right ascension and declination of the galaxy as listed in the \citet{vanderwel12} F160W catalog, with the same position angle as the best-fit \texttt{GALFIT} model in \citet{vanderwel12}.  We do not calculate half-mass radii for galaxies that do not have \sersic models listed in the \citet{vanderwel12} catalog. We fix the axis ratio of the annuli to a convolved version of the \texttt{GALFIT} axis ratio, $q_{\rm{conv}} = \sqrt{(r_e^2 + r_{PSF}^2) / ([q r_e]^2 + r_{PSF}^2)}$, where $r_{PSF}$ is the half width at half maximum (HWHM) of the PSF and $r_e$ is effective radius measured by GALFIT in the F160W band. Using this convolved axis ratio allows us to extract photometry that more closely follows the geometry of the observed galaxy. 
The semimajor axis of each annulus is set to an integer value of $r_\text{PSF}$; this ensures that we are not measuring photometry on scales that are entirely ``blurred out'' by PSF effects. Because all images are convolved to the F160W PSF with a FWHM of 0.19 arcseconds, $r_\text{PSF}$ is equal to 0\farcs095 ($\sim0.5$~kpc over our full redshift range).

We use the {\it photutils} python package to measure aperture photometry in each annulus, repeating the measurement for each galaxy in all available 3D-HST PSF-matched images. For galaxies in the COSMOS and UDS fields, this typically yields five bands in our resolved SED: F160W, F125W, F140W, F606W, and F814W. For galaxies in the GOODS-S field, there are typically eight bands in our resolved SED: F160W, F125W, F140W, F435W, F606W, F775W, F850LP, and F814W. 
Errors on the measured aperture photometry are calculated using the empty aperture scaling law described in \citet{skelton14}, which parameterizes the error on a photometric measurement in terms of the area of the measurement aperture and the value of the weight map in the aperture. 
We define the outermost annulus to be the last annulus where the signal-to-noise ratio of the photometry in the F160W band is greater than 10.0.  
Beyond this radius, the low S/N of the photometry in each annulus does not allow for a robust mass determination. Each galaxy has measured aperture photometry for 2-43 annuli; most galaxies have photometry in 7-10 annuli. When calculating aperture photometry, we do not include flux from pixels that are identified as part of another adjacent galaxy in the segmentation maps.  

Next, we use the SPS fitting code FAST \citep{kriek09} to independently fit the SED of each annulus. In these fits, we assume the \citet{bc03} stellar population library, a \citet{chabrier03} initial mass function, a delayed exponential star formation history, and the \citet{kriek13} dust attenuation law. Fitting each annulus separately allows the mass, age, star-formation timescale, and dust extinction of each annulus to vary independent of the best-fit values for other annuli and the best-fit values for the galaxy as a whole. The free parameters in the fit are the age, star formation timescale $\tau$, and dust extinction $A_v$ of the annulus. The mass is not a free parameter: it is set by the overall scaling of the measured photometry. We fix the redshift of each annulus to the ZFOURGE photometric (or spectroscopic, when available) redshift of the galaxy so that different regions of the galaxy are not modeled at different best-fit redshifts. We allow $\log(\tau{\rm{/yr}})$ to vary between 7.0 and 10.0 in steps of 0.2, $\log({\rm{age/yr}})$ to vary between 8.0 and 10.0 in steps of 0.1, and $A_v$ to vary between 0.0 and 3.0~mag in steps of 0.1~mag; the metallicity is fixed to $0.02\ Z_\odot$. We use the FAST version 0.2 template error function.

By considering only the $5-8$ filters where we can measure spatially resolved photometry, we are ignoring the wealth of ancillary data available in these well-studied extragalactic fields. In addition to the resolved photometry, each galaxy has integrated photometry in $\sim 20$ additional filters ranging from the UV to near-IR. Following \citet{wuyts12}, we use this integrated photometry to adjust the best-fit model for each annulus so that the sum of the modeled SEDs for each annulus matches the observed photometry of the galaxy as a whole. 
Mathematically, this amounts to minimizing a $\chi^2$ equation with two terms. The first term describes how well the SPS model for each annulus fits the resolved photometry in that annulus. This first term is what SPS fitting codes like FAST minimize: $$\chi^2_{\text{res}} = \sum_{i=1}^{N_{\text{annuli}}} \sum_{j=1}^{N_{\text{res}}}\frac{(F_{i, j} - M_{i, j})^2}{(E_{i,j})^2},$$
where $N_{\text{annuli}}$ is the total number of elliptical annuli, $N_{\text{res}}$ is the number of filters with spatially resolved photometry, $F_{i,j}$ represents the measured flux in annulus $i$ and filter $j$, $M_{i,j}$ represents the modeled flux in annulus $i$ and filter $j$, and $E_{i,j}$ is the error on the measured flux for annulus $i$ and filter $j$.

The second term in our $\chi^2$ function describes how well the sum of the modeled SEDs of all annuli matches the observed integrated photometry for the entire galaxy: $$\chi^2_{\text{int}} = \sum_{j=1}^{N_{\text{int}}}\frac{(F_j - \sum_{i=1}^{N_{\text{annuli}}}M_{i,j})^2}{(E_j)^2}.$$ Here, $N_{\text{int}}$ is the number of bands with integrated photometry, $M_{i,j}$ again represents the modeled flux in annulus $i$ and filter $j$, and $F_j$ represents the observed integrated photometry in filter $j$. To account for aperture differences between the catalog and our largest annulus, we scale the integrated photometry down by the error-weighted average of the difference between the catalog flux and sum of the annuli fluxes in all bands with resolved photometry. This correction factor is typically between $\sim 0.75$ and $1$. 

Unlike \citet{wuyts12}, who minimize $\chi^2_{\text{res}} + \chi^2_{\text{int}}$, we minimize $$\chi^2_{\text{tot}} = \frac{1}{\nu_{\text{res}}} \chi^2_{\text{res}} + \frac{1}{\nu_{\text{int}}} \chi^2_{\text{int}},$$ where $\nu$ is the number of bands fitted minus the number of free parameters in the fit (three: age, $\tau$, and $A_v$). We account for the number of bands we fit in our total $\chi^2$ because there are many more bands with integrated photometry than resolved photometry; without this term, the minimization essentially ignores how well the model for each annulus fits the resolved photometry in that annulus in favor of ensuring that the sum of the annuli models exactly matches the integrated photometry. This could result in the individual annuli fits, which we ultimately use to measure the mass profiles, to be quite poor.

Following \citet{wuyts12}, we use an iterative approach to minimize $\chi^2_{\text{tot}}$. We use the best-fit values from the resolved photometry SED modeling for the age, $\tau$, and $A_v$ of each annulus as an initial condition, and evaluate $\chi^2_{\text{tot}}$. Then, we allow the age, $\tau$, and $A_v$ of the innermost annulus to simultaneously move to an adjacent position on the FAST grid and re-calculate $\chi^2_\text{tot}$ for each of the possible new (age, $\tau$, $A_v$) combinations. The age, $\tau$, and $A_v$ for the innermost annulus that yield the smallest $\chi^2_{\text{tot}}$ are taken to be the new best-fit parameters for that annulus. We repeat this process of finding the new best-fit values for each of the remaining annuli in turn. We continue to repeat this process of finding new best-fit values for all annuli until $\chi^2_{\text{tot}}$ remains constant or we reach 500 iterations; in practice, most galaxies converge in $20-100$ iterations. At the end of this constraint process, we have determined the best-fit age, $\tau$, and $A_v$ for each annulus--- including information from the integrated bands--- and can read off the best-fit mass that corresponds to these best-fit parameters for each annulus. 

We use a Monte Carlo technique to determine error bars on the mass of each annulus. We use the FAST 1$\sigma$ error contour to randomly select a new best-fit starting age, $\tau$, and $A_v$ for each annulus. We repeat the constraint process beginning from these perturbed initial conditions. Additionally, we vary the total flux values for the integrated filters according to the error bars listed in the catalog. We perform 200 of these simulations, and take the 68\% confidence interval on the resulting annuli masses as our $1\sigma$ error bars. To ensure that these error bars are realistic, we enforce a minimum 10\% error on the annuli masses.

To obtain a mass-to-light ratio measurement, we use EAZY to interpolate the observed aperture photometry to a rest-frame $g$ band profile, assuming the SDSS $g$-band filter curve. We then report mass-to-light ratios as $M/L_g$. We also report $M/L_{F160W}$, the mass-to-light ratio in the {\it observed} F160W filter; these measurements are used in conjunction with morphologies measured from F160W images. We stress that, at this point in our analysis, the mass and $M/L$ profiles are based on the observed (convolved-space) data, mitigating the large uncertainties that can arise when using deconvolved data.

Despite the advantages of measuring the mass profile in convolved space--- namely, that the measured profile does not strongly depend on the deconvolution algorithm--- measurements in convolved space can only tell us whether the half-mass radii of these galaxies are larger, smaller, or the same as their half-light radii. In this paper, we also wish to quantify {\it how much} larger or smaller the half-mass radii are than the half-light radii. This requires correcting for the PSF and moving to deconvolved space.
In this paper, we use two different techniques to recover intrinsic $M/L$ profiles and half-mass radii from the observed $M/L$ gradients we find from spatially-resolved SED fitting. 

\subsection{Method 1: interpreting observed M/L gradients with a simple forward modeling technique}
\label{sec:method1}

We first test a simple forward modeling technique to fit the observed $M/L_{F160W}$ profile: we make a series of models assuming different intrinsic $M/L$ profiles, convolve each model with the PSF, then use $\chi^2$ minimization to find the best intrinsic $M/L$ profile and half-mass radius. Details of this method are described below.

We assume that the intrinsic 2D light profile of the galaxy follows the \texttt{GALFIT}-derived best-fit S\'ersic model measured in \citet{vanderwel12} using the F160W CANDELS images. We evaluate this S\'ersic model on a 2D grid with the same pixel scale as the data. Then, we apply a range of possible intrinsic $M/L$ profiles to obtain a series of possible intrinsic mass profiles for the galaxy. We parameterize the intrinsic $M/L$ profile as a power-law function of radius $r$--- i.e., $\log{M/L} \propto \alpha \times \log{r}$ --- as suggested by, e.g., \citet{chan16}. The slope of the relation is allowed to vary between $-2.3 \le \log{([M/L|_{2r_{\text{e,light}}}] / [M/L|_{r_{\text{e,light}}}])} \le 2.3$; these bounds were chosen to fit the range of observed convolved $M/L$ profiles in this sample. It is clearly possible for galaxies to have more complex radial $M/L$ profiles than a simple power law; however, many of our galaxies are not significantly larger than the PSF and we do not typically have a large number of data points in our $M/L$ profiles. Including a large number of parameters in our intrinsic $M/L$ model would quickly result in overfitting. 

We set the $M/L$ profile at radii smaller than one pixel (0\farcs06) to the value at one pixel, and set the $M/L$ profile at very large radii to the last value where we have measured aperture photometry. This prevents artificially small (large) half-mass radii from strongly decreasing (increasing) $M/L$ profiles.  

We then convolve both the 2D light profile and all possible 2D mass profiles with the F160W PSF  (using the F160W PSF for each field as provided by the 3D-HST team). We extract the 1D convolved mass and light profiles in annuli, again using the {\it photutils} python package to measure aperture photometry. The resulting model $M/L_{F160W}$ profiles are arbitrarily normalized to $M/L=1$ at $r = r_\text{e, light}$. We scale the model $M/L_{F160W}$ profiles to the observed $M/L_{F160W}$ profile by multiplying the model by $\Sigma(\text{model} \times \text{data}) / \Sigma(\text{model}^2)$. Then, we calculate the $\chi^2$ value between the modeled and observed $M/L$ profiles and take the model with the smallest $\chi^2$ as the best fit. If there is no model with $\chi^2 < 15$, we do not report a half-mass radius for that galaxy.

We calculate the half-mass radius for the best-fit intrinsic mass model using the analytic 1D S\'ersic profile (based on the F160W \texttt{GALFIT} model) instead of the `pixelated' 2D profile to ensure accuracy. We multiply this analytic 1D S\'ersic profile by the best-fit intrinsic $M/L$ profile, and evaluate the profile out to 500 pixels (30") to ensure that we have captured the total mass of the galaxy. We then find the half-mass radius by calculating the radius where the mass profile reaches half its total value. 

We use a Monte Carlo method to estimate the uncertainties of the best-fit intrinsic $M/L$ model. We choose a new half-light radius and S\'ersic index according to the error bars listed in \citet{vanderwel12}, and create a new grid of models. We perturb the observed $M/L$ profile according to its error bars, then use $\chi^2$ minimization to find the best-fit model. We record the 68\% confidence intervals of 200 such simulations as the 1$\sigma$ error bars on the half-mass radius and $M/L$ slope of the models.  

Figure \ref{fig:method1} shows example images, SEDs, $M/L$ profiles, and mass profiles derived using this method for three galaxies in our sample. Appendix \ref{sec:recovery} shows that the half-mass radii recovered using this method do not have significant biases with stellar mass, half-light radius,  redshift, or the number of annuli in the fit.

\subsection{Method 2: interpreting observed M/L gradients with \texttt{GALFIT}}
We also test a second way to derive half-mass radii and intrinsic $M/L$ gradients from the $M/L$ profiles measured in Section \ref{mainmethod}. This recovery technique is similar to that used by \citet{lang14} and \citet{chan16}: we create an as-observed 2D mass map for each galaxy, then fit it with \texttt{GALFIT} to find the intrinsic half-mass radius. We describe this method in more detail below.

To create a mass map, we first smooth the 1D observed $\log{M/L} - \log{r_{\rm{obs}}}$ profile using a cubic spline interpolation; this decreases the edge effects at the boundaries between annuli and results in better residuals on the \texttt{GALFIT} fits. As with the forward-modeling method described above, we fix $M/L$ at radii larger than $r_{\text{max}}$ to the value at $r_{\text{max}}$. Then, we map this profile from 1D to 2D using the ellipticity and position angle of the elliptical annuli. Multiplying this 2D $M/L_{F160W}$ map by the original F160W image of the galaxy yields a mass map that we then fit with \texttt{GALFIT}. 

Unlike \citet{chan16}, we keep the same range of allowed \sersic indices in both the mass and light profile fits. Because the best-fit \sersic indices for the light profiles of many compact quiescent galaxies reaches the maximum value of $n=8.0$, allowing a different range of $n$ values between the light and mass \texttt{GALFIT} fits may result in measurements of the half-light radii being biased with respect to measurements of the half-mass radii. Following \citet{vanderwel12}, then, in our \texttt{GALFIT} fits to the mass maps we allow $n$ to vary between 0.2 and 8.0 and $r_e$ to vary between 0.3 and 400.0 pixels. We do not fix the total mass, position, position angle, or axis ratio of the best-fit model. Because the F160W images are already sky-background subtracted, we fix the sky background to zero in the fits.

We again turn to a Monte Carlo method to determine error bars on the measured half-mass radii. We perturb each $M/L$ measurement within its error bars, re-create a mass map, and re-fit it with \texttt{GALFIT}. We take the 68\% confidence interval of 200 such realizations as our $1\sigma$ error bars. The error bars are dominated by uncertainties in our $M/L$ profile, and are significantly larger than the errors estimated by \texttt{GALFIT}. 

\subsection{Method 3: The Szomoru et al. method}
For each galaxy, we also derive mass profiles and half-mass radii following the method used by \citet{szomoru10,szomoru12,szomoru13}. This method relies on intrinsic surface brightness profiles, which are used to measure a rest-frame $u-g$ color; an empirical relation between this color and the mass-to-light ratio is then used to derive a mass profile. We describe this method in detail below.

We fit each galaxy in all available filters using \texttt{GALFIT} \citep{peng02}. The input image used is a $\sim3"$ cutout of the 3D-HST image in a given filter, centered at the galaxy's position as listed in the \citet{vanderwel12} catalog. As for the two methods described above, we use the images convolved to the F160W PSF; this ensures that we perform a fair comparison between the the three methods by beginning with identical data products. The `sigma' image is constructed as the inverse square root of the 3D-HST weight image. We use the `mask' image to exclude all pixels identified in the 3D-HST segmentation map as belonging to adjacent sources. Because all images are convolved to the F160W PSF, we use the F160W PSF as provided by 3D-HST as the input PSF for \texttt{GALFIT}. Following \citet{vanderwel12}, we constrain the \sersic index $n$ of the best fit to lie between 0.2 and 8.0, the effective radius $r_e$ to lie between 0.3 and 400.0 pixels, and the total magnitude of the best-fit model to be within three magnitudes of the galaxy's catalog magnitude in the filter being fit.

If \texttt{GALFIT} ran successfully --- i.e., it did not reach the maximum number of iterations and there were no `suspected numerical convergence errors' --- we construct the galaxy's surface brightness profile. We measure the flux in the \texttt{GALFIT} residual image in concentric annuli with the same geometry as the best-fit model, then add these residuals to the best-fit \sersic model. Following \citet{szomoru12}, we correct the best-fit \sersic model with residuals only out to 10kpc; the profile from 10-100kpc is not residual-corrected because of difficulties in constraining the sky background at these scales. We derive errors on the intrinsic profiles using the empty aperture scaling laws described in \citet{skelton14}.  After ensuring that we have measured surface-brightness profiles both redward and blueward of the SDSS $u$ and $g$ filters, we use \texttt{EAZY} \citep{brammer08} to interpolate the observed surface brightness profiles into rest-frame $u$ and $g$ profiles. Because of this interpolation, half-mass radii at $z > 2.23$-- where the central wavelength of the redshifted $g$ band is redder than the central wavelength of the F160W filter-- cannot be measured using this technique.

A galaxy's rest-frame $u-g$ color is well correlated with its mass-to-light ratio, $\log M/L_g$, primarily because changes in stellar age, metallicity, and dust attenuation are degenerate in the $u-g$ - $\log M/L_g$ space (Bell \& de Jong 2001). We use the masses and rest-frame colors for the full 3D-HST survey to determine the best-fit linear relation between $u-g$ and $\log M/L_g$. Because the slope of the relation varies as a function of redshift (Szomoru 2013), we derive the best-fit relation for each target galaxy in a redshift slice of width $\Delta z=0.4$ centered at the target's redshift. We use this best-fit relation to calculate the M/L$_g$ profile of each galaxy in our sample. We then multiply the $\log M/L_g$ profile by the \texttt{EAZY}-derived $L_g$ profile to obtain the mass profile of the galaxy. Finally, the half-mass radius is calculated as the radius where the mass profile reaches half of its maximum value. 

Errors on the half-mass radius are estimated using a Monte Carlo technique: we vary the $u-g$ profile within its 1$\sigma$ error bars, re-derive an M/L$_g$ profile, and re-calculate the half-mass radius. In contrast to \citet{szomoru10,szomoru12,szomoru13}, we also include uncertainties caused by scatter in the $u-g$ - $\log\rm{M}/\rm{L}_g$ relation by perturbing each simulated $\log\rm{M}/\rm{L}_g$ profile according to the observed scatter in the $u-g$ - $\log\rm{M}/\rm{L}_g$ relation.

%------------------------------------------------------------------------------------------------------------------------------------------------------------------------
\section{Results: half-mass radii for $\sim7,000$ galaxies}

\begin{table*}
\caption{Half-mass radii for 7,006 galaxies.}
\begin{threeparttable}
\begin{tabular}{cccccccccccc}
\hline \hline
ID\tnote{a} & Field & RA\tnote{b} & Dec\tnote{b} & z\tnote{c} & $\log{\frac{M_*}{M_\odot}}$\tnote{d} & $r_{\rm{e,light}}$ (kpc)\tnote{e} & n\tnote{f} & \shortstack{$r_{\rm{e,mass}}$ (kpc) \\ Method 1} & \shortstack{$r_{\rm{e,mass}}$ (kpc) \\ Method 2} & \shortstack{$r_{\rm{e,mass}}$ (kpc) \\ Method 3} & \\
\hline
    10571 & GOODS-S & 53.101357 & -27.860300 & 2.04 & 9.45 & $0.603 \pm 0.056$ & $3.11 \pm 0.67$ & $0.660^{+1.071} _{-0.631}$ &  - & $0.827^{+0.998} _{-0.706}$ & \\
    10784 & GOODS-S & 53.130981 & -27.860407 & 1.37 & 9.99 & $3.143 \pm 0.073$ & $3.75 \pm 0.11$ & $2.064^{+2.104} _{-1.967}$ & $2.413^{+2.392} _{-2.280}$ & $2.829^{+3.219} _{-2.586}$ & \\
    13870 & GOODS-S & 53.083248 & -27.847992 & 1.27 & 10.06 & $5.560 \pm 0.052$ & $0.68 \pm 0.02$ & $4.400^{+4.448} _{-4.354}$ & $4.103^{+4.110} _{-4.100}$ & $5.006^{+5.318} _{-4.727}$ & \\
    14567 & GOODS-S & 53.089684 & -27.844601 & 1.49 & 9.99 & $5.045 \pm 0.066$ & $0.88 \pm 0.04$ & $3.474^{+3.553} _{-3.414}$ & $3.377^{+3.459} _{-3.366}$ & $4.030^{+4.428} _{-3.704}$ & \\
    17184 & GOODS-S & 53.123787 & -27.832561 & 1.04 & 9.09 & $2.399 \pm 0.119$ & $1.32 \pm 0.18$ & $1.856^{+2.003} _{-1.744}$ & $1.898^{+1.948} _{-1.893}$ & $2.572^{+3.216} _{-2.229}$ & \\
    17469 & GOODS-S & 53.190910 & -27.831028 & 1.16 & 9.47 & $2.006 \pm 0.020$ & $0.52 \pm 0.02$ &  - & $1.757^{+1.758} _{-1.757}$ & $2.117^{+2.287} _{-1.979}$ & \\
    18116 & GOODS-S & 53.186100 & -27.827543 & 1.98 & 9.95 & $3.846 \pm 0.158$ & $2.48 \pm 0.14$ & $3.437^{+3.617} _{-3.268}$ & $2.900^{+2.977} _{-2.684}$ & $1.941^{+2.287} _{-1.683}$ & \\
    19865 & GOODS-S & 53.081169 & -27.818588 & 1.24 & 9.55 & $2.525 \pm 0.038$ & $1.02 \pm 0.04$ & $2.190^{+2.204} _{-2.086}$ & $2.064^{+2.143} _{-2.061}$ & $2.695^{+3.031} _{-2.489}$ & \\
    21868 & GOODS-S & 53.226711 & -27.808552 & 1.91 & 9.82 & $1.642 \pm 0.041$ & $0.78 \pm 0.09$ & $1.621^{+1.711} _{-1.358}$ & $1.920^{+1.899} _{-1.581}$ & $1.557^{+1.805} _{-1.367}$ & \\
    ... & ... &... &... &... &... &... &... &... &... &... & \\
    \hline
\end{tabular}
\begin{tablenotes}
\item[a]{From the v4.0 3D-HST catalog.}
\item[b]{Taken from \citet{vanderwel12} catalog to match the morphological measurements.}
\item[c]{$z_p$ taken from the ZFOURGE catalog \citep{straatman16}. By comparing photometric and spectroscopic redshifts, \citet{straatman16} estimate photometric redshift errors of $\Delta z / (1+z) \sim 0.01$.}
\item[d]{Taken from the ZFOURGE catalog \citep{straatman16}. Corrected to be consistent with morphological measurements by multiplying the catalog mass by the ratio of the total F160W flux measured by \texttt{GALFIT} to the total F160W flux measured in the \citet{straatman16} catalog. Uncertainties on stellar masses are dominated by systematics, and are estimated to be $\sim 0.2 - 0.3$~dex}.
\item[e]{From the \citet{vanderwel12} catalogs. Corrected to rest-frame 5,000\AA\ using the procedure in \citet{vanderwel14}.}
\item[f]{From the \citet{vanderwel12} catalogs.}
\item{All listed radii represent measurements of the {\it major axis} of the galaxy.}
\item (This table is available in its entirety in a machine-readable form in the online journal. A portion is shown here for guidance regarding its form and content.)
\end{tablenotes}
\end{threeparttable}
\label{tab:results}
\end{table*}

Table \ref{tab:results} lists the half-mass radii for all galaxies in the sample calculated using each of the three methods described above. Other basic galaxy properties, such as stellar mass and redshift, are also listed. The full table is available online. The total number of galaxies successfully fit with each method differs. Method 1 succeeded for 5,649 galaxies-- the rest were not well-fit with our power-law model for $M/L$. Method 2 succeeded for 6,552 galaxies; the rest did not have successful \texttt{GALFIT} mass profile fits. Method 3 succeeded for 6,072 galaxies; the rest did not have measured surface brightness profiles both redward and blueward of the rest-frame $u$ and $g$ filters.

\begin{figure*}
    \centering
    \includegraphics[width=\textwidth]{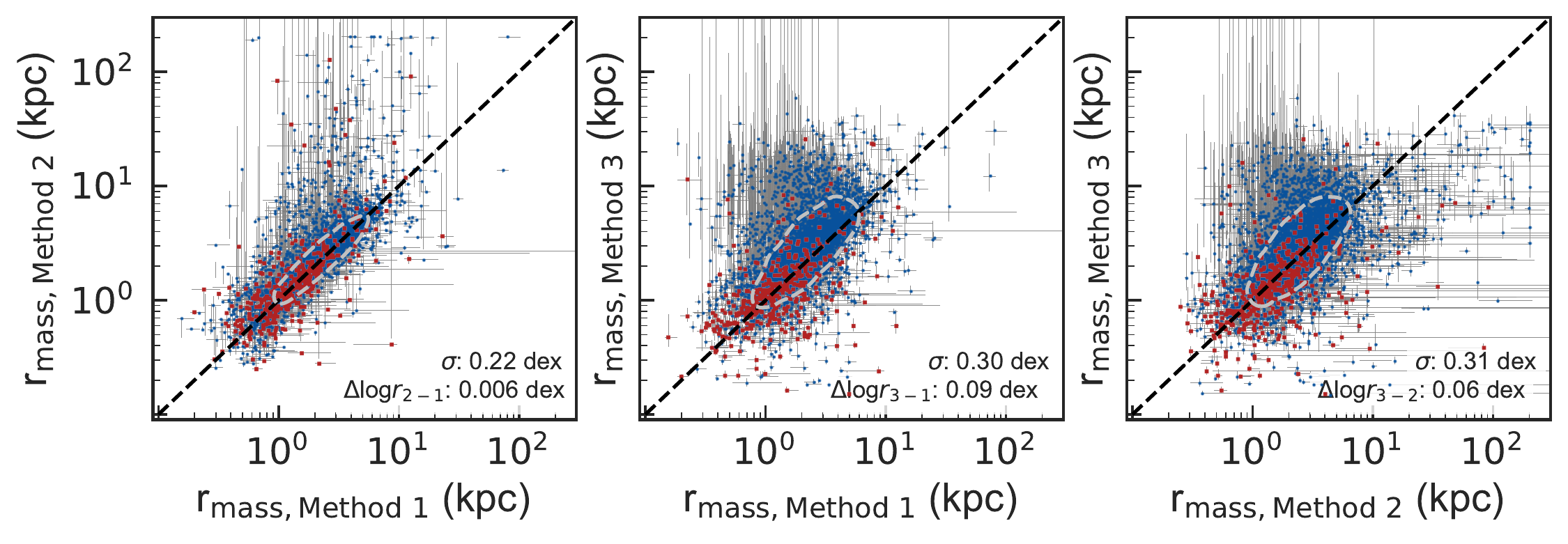}
    \caption{Comparison of half-mass radii calculated with three different methods (described in Section \ref{sec:methods}). Red squares represent galaxies classified as quiescent using a UVJ selection; blue points represent star-forming galaxies in a UVJ selection. The grey dashed ellipse represents the 1$\sigma$ error contour, and the black dashed line shows the one-to-one relation.}
    \label{fig:rvr}
\end{figure*}

Figure \ref{fig:rvr} compares the half-mass radii of all galaxies in our sample calculated using each of the three methods discussed in Section \ref{sec:methods}. UVJ-classified star-forming galaxies are shown in blue, and quiescent galaxies are shown in red. The black dashed line shows the one-to-one relation, and the grey dashed ellipse shows the middle 68\% of the points. On average, we see that galaxy half-mass radii are remarkably consistent for all three measurement techniques. 

As expected, Methods 1 \& 2 (which fit the same observed M/L gradient, but use different techniques to account for the PSF), produce very consistent results. There is a typical scatter of 0.22~dex between the two measurements; the median offset between the two sets of half-mass radii is very small, with $\Delta r_{\rm{mass}} < 0.01$~dex. 

There is slightly more scatter, $\sim$0.3~dex, when comparing half-mass radii calculated using Method 3 to those calculated using Method 1 or 2. The Method 3 half-mass radii are also offset slightly larger than the Method 1 \& 2 half-mass radii, by 0.09~dex and 0.06~dex respectively. This systematic bias is especially apparent for large star-forming galaxies. Intriguingly, this is similar to the offset that \citet{price17} found between the intrinsic half-mass radii of simulated massive galaxies and the half-mass radii recovered using their implementation of the \citet{szomoru10,szomoru12,szomoru13} method (our Method 3).
 
 The simplest explanation for the offset between the Method 3 radii when compared to the Method 1 \& 2 radii is that imperfect sky modeling at large radii led to errors in the deconvolved Method 3 profiles. If the sky subtraction is not perfect at large radii, then the shape of each \texttt{GALFIT} profile at large radii is incorrect. Both the $u$ and $g$ profiles will converge to approximately the same value, the noise level. Because the outskirts of most galaxies tend to be bluer than the centers (explored further in Section \ref{sec:trends}), this would lead to the observed color gradient being {\it flatter} than the true, blue, gradient. Consequently, $M/L$ would be overestimated, $r_{\rm{mass}}/r_{\rm{light}}$ would be overestimated, and the half-mass radius for Method 3 would be overestimated. This effect would be more apparent for large galaxies, as more of their outskirts would be subsumed by residual sky noise. Since blue galaxies are typically larger than red galaxies at fixed stellar mass, this effect is also generally more apparent for blue galaxies. Again, we note that the magnitude of the difference in half-mass radii calculated with Method 3 as opposed to Method 1 or 2 is only $\sim$0.3~dex. 
 
 In Appendix \ref{sec:methodtest}, we further investigate how well the half-mass radii measured using Methods 1 and 3 agree. We show that the differences between the two sets of half-mass radii do not depend on galaxy stellar mass, redshift, \sersic index, or $U-V$ color.

Overall, the consistency between half-mass radii calculated with Methods 1, 2, \& 3 indicates that, despite difficulties in measuring half-mass radii and in accounting for the PSF, these methods are converging towards a physically-meaningful half-mass radius. 
While our conclusions for the remainder of the paper do not depend on the method we use to calculate half-mass radii, we wish to pick one method as the `primary' method to show in our plots.

Each of the three methods has its own strengths and weaknesses. Using both Methods 1 \& 2, we can examine the observed-space $M/L$ profile before performing modeling to account for the PSF. This allows us to verify whether half-mass radii are smaller, larger, or equal to half-light radii with minimal dependence on modeling. Furthermore, because these methods model the full SED of each annulus, they make use of all of the available multi-band data. Method 1 uses a fairly simple and easy-to-interpret approach to account for PSF effects; however, complex M/L profile shapes are likely not well-described by our power law function. Method 2-- which uses GALFIT to fit a mass map-- suffers from issues common to this type of modeling: it can be difficult to perfect sky subtraction at large radii, and galaxies with non-\sersic mass distributions are not modeled well. Method 3 is quite straightforward and easy-to-implement. It is also ideal for cases where only a few bands of imaging are available, because it is based off of a single color; by the same token, this method does not fully utilize the multi-band data available for the galaxies in our sample. Furthermore, by accounting for PSF effects in each band separately and then subtracting deconvolved profiles, any possible errors in the deconvolution process can have a large effect on the final half-mass radius. Considering these strengths and weaknesses, we choose to present our results below using the Method 1 half-mass radii. Again, we stress that our conclusions are unchanged if we instead use the half-mass radii from Method 2 or 3.

\subsection{Quantifying the strength of galaxy color gradients}
\label{sec:trends}
The ratio of half-mass to half-light size, ${r_{\rm{mass}} / r_{\rm{light}}}$, describes the strength of the color gradient in a galaxy. If there is no color gradient present, the half-mass and half-light sizes will be equal. If there is a weak negative color gradient, meaning that the galaxy is slightly bluer at larger radii, the half-mass size will be slightly smaller than the half-light size and ${r_{\rm{mass}} / r_{\rm{light}}} < 1$; a strong negative color gradient will result in a much smaller half-mass size than half-light size, and ${r_{\rm{mass}} / r_{\rm{light}}} \ll 1$. Similarly, a weakly positive color gradient means that ${r_{\rm{mass}} / r_{\rm{light}}} > 1$ and a strong positive color gradient means that ${r_{\rm{mass}} / r_{\rm{light}}} \gg 1$. By examining ${r_{\rm{mass}} / r_{\rm{light}}}$, we can examine the strength of color gradients in galaxies and how they correlate with other galaxy properties. 

\begin{figure*}
    \centering 
    \includegraphics[width=\textwidth]{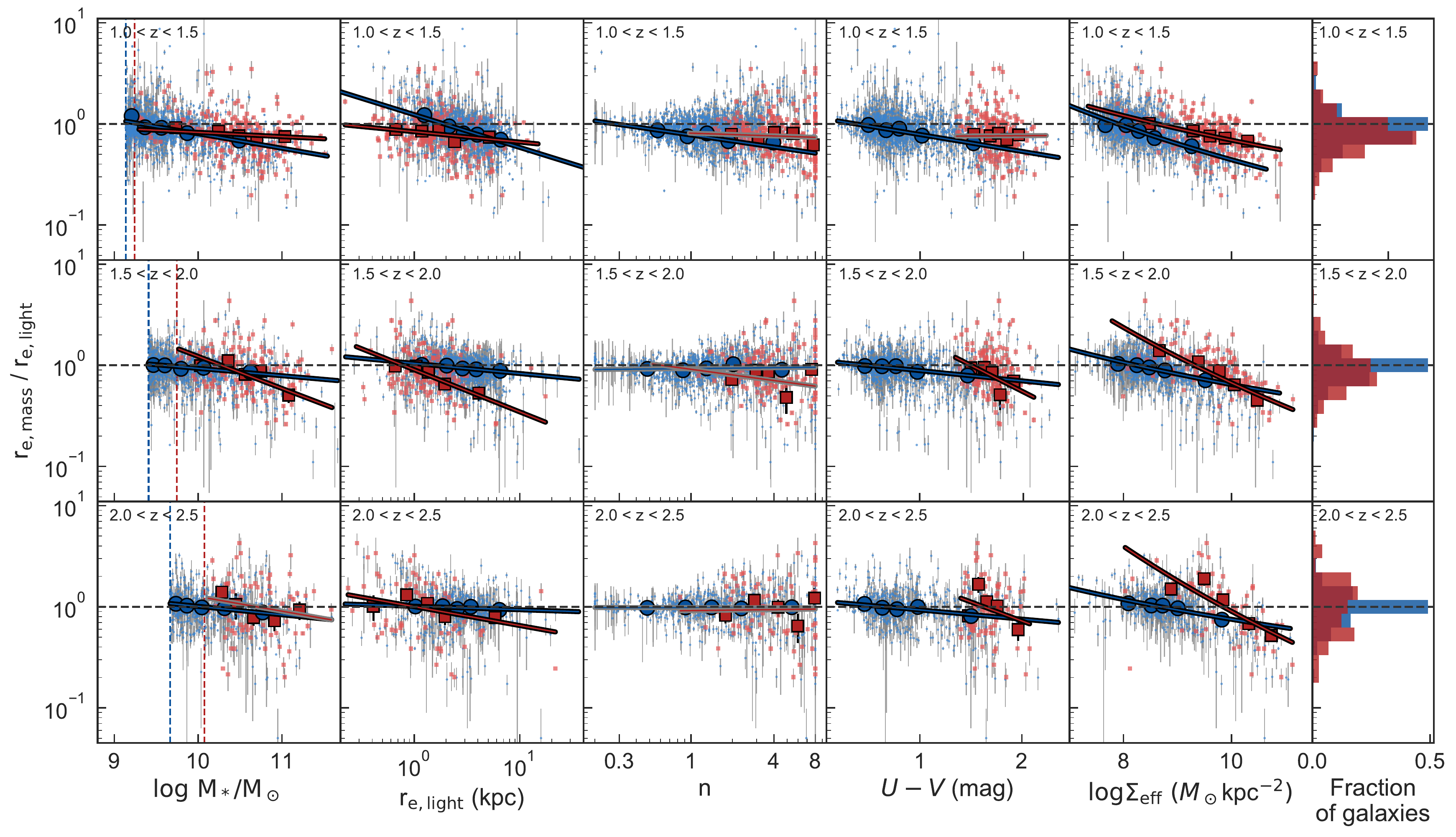}
    \caption{Half-mass to half-light ratio as a function of other galaxy properties. Each row shows one redshift slice. From left to right, each column shows the correlation with: stellar mass, half-light radius, \sersic index, $U-V$ color, and mass surface density $\Sigma_{\rm{eff}} = 0.5 M_* / (\pi r_{\rm{e,mass}}^2)$. The rightmost column shows a histogram of $r_{\rm{e,mass}} / r_{\rm{e,light}}$ for all star-forming (blue) and quiescent (red) galaxies in the redshift range. Individual star-forming and quiescent galaxies are plotted as small light blue circles and small light red squares. The running error-weighted mean of star-forming and quiescent galaxies is plotted as large blue circles and large red squares; error bars on these mean points are the standard error on the mean ($\sigma / \sqrt{N})$. The data set is binned such that each mean point represents the same number of individual data points. Best fits to the trends for star-forming and quiescent galaxies are shown as solid blue and red lines. Best-fit lines are outlined in black if the slopes are inconsistent with zero, and outlined in grey if the slopes are consistent with zero. The horizontal grey dashed line in each panel indicates where $r_{\rm{mass}} = r_{\rm{light}}$. Mass completeness limits for star-forming and quiescent galaxies are shown as blue and red dashed vertical lines in the leftmost column; galaxies below the mass completeness limits are neither included in the fits nor plotted in this figure.}
    \label{fig:trends}
\end{figure*}

In Figure \ref{fig:trends}, we show correlations between ${r_{\rm{mass}} / r_{\rm{light}}}$ and galaxy stellar mass, half-light radius, the \sersic index $n$ measured from the galaxy's F160W light profile, $U-V$ color, and stellar mass surface density within the {\it circularized} effective radius $\Sigma_{\rm{eff}} = 0.5 M_* / (\pi [r_{\rm{e, mass}} \times{\sqrt{b/a}}]^2)$. We use a circularized half-mass radius to compute $\Sigma_{\rm{eff}}$ for consistency with other works. To minimize the effects of any redshift evolution on the observed correlations, we show three different redshift intervals. We only plot galaxies with masses above the mass completeness level for each redshift interval (dashed vertical lines in left column of Figure \ref{fig:trends}). Small blue points and red squares show individual star-forming and quiescent galaxies, and large blue points and red squares show a running error-weighted average. 

\begin{figure}
    \centering
    \includegraphics[width=.45\textwidth]{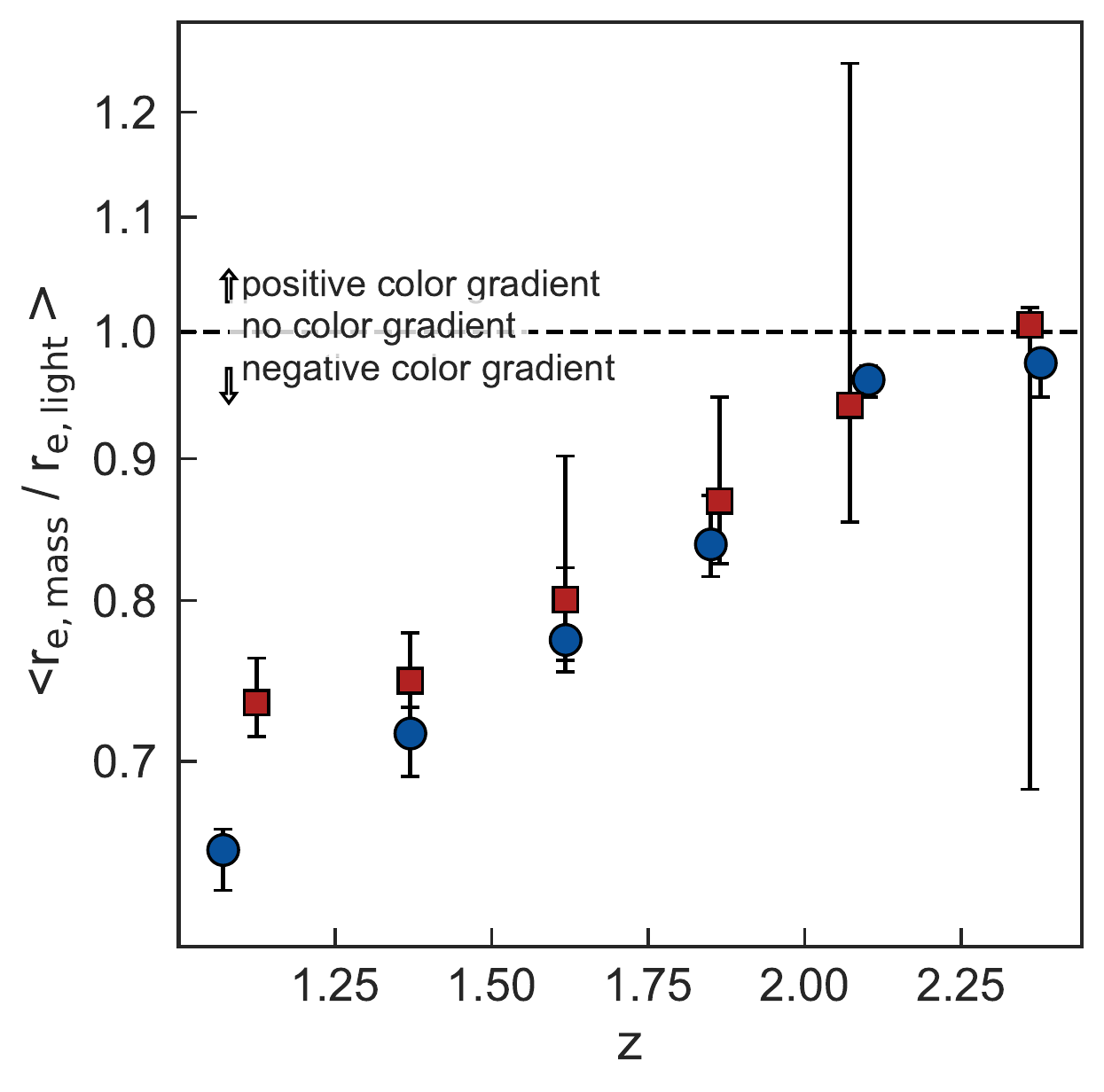}
    \caption{Median half mass-to-half light ratio for bins of star-forming (blue circles) and quiescent (red squares) galaxies as a function of median redshift. Only galaxies with masses greater than the most stringent mass completeness cut ($\sim 10^{10} M_\odot$) are included. Error bars represent the middle 68\% of 500 bootstrap samples.}
    \label{fig:ratioZtrend}
\end{figure}

We use a least squares technique to fit a line of the form $\log{r_{\rm{mass}} / r_{\rm{light}}} = s (\log{M/M_\odot} - 10) + b$ to the trend as a function of mass, a line of the form $\log{r_{\rm{mass}} / r_{\rm{light}}} = s (\log{r_{\rm{e,light}}/{\rm{kpc}}} - 1) + b$ to the trend as a function of half-light radius, a line of the form $\log{r_{\rm{mass}} / r_{\rm{light}}} = s (\log n - 1) + b$ to the trend as a function of \sersic index, a line of the form $\log{r_{\rm{mass}} / r_{\rm{light}}} = s ([U-V] - 1) + b$ to the trend as a function of $U-V$ color, and a line of the form $(\log{r_{\rm{mass}} / r_{\rm{light}}} = s (\log{\Sigma_{\rm{eff}}} - 9) + b)$ to the trend as a function of stellar mass surface density. We fit the individual data points, not the mean points also shown in Figure \ref{fig:trends}.
We estimate error bars on the fits by performing 500 bootstrap simulations. The best-fit slopes and intercepts for these fits, as well as their error bars, are listed in Table \ref{table:line_fits}. The best-fit relations for all parameters are also plotted in Figure \ref{fig:trends}. The fits are outlined in grey if their slopes are consistent with zero within the $1\sigma$ error bars, and outlined in black if the slopes are not consistent with zero. We see significant trends in $r_{\rm{mass}} / r_{\rm{light}}$ with both stellar mass, stellar mass surface density, half-light radius, $U-V$ color, and $\Sigma_{\rm{eff}}$ for blue and red galaxies. In the highest-redshift bin, the trends are generally less significant. This is likely due to flatter color gradients in this redshift interval (explored further in Figure \ref{fig:ratioZtrend}). The strongest trend we see is in $r_{\rm{mass}} / r_{\rm{light}}$ as a function of $\Sigma_{\rm{eff}}$. Interestingly, this trend remains significant in the highest-redshift bin, even though trends in other galaxy properties are often consistent with being flat.

To ensure that the trends we see are not driven by differences in mass completeness cuts for star-forming and quiescent galaxies at different redshifts, we also calculated the trends shown in Figure \ref{fig:trends} while considering only galaxies with $M_*>10^{10.08}M_\odot$. This mass corresponds to the most stringent mass completeness cut in our sample. We saw no significant differences in the best-fit relations, indicating that these trends are not driven by the exact value of our mass completeness limits.

\begin{table}[]
\centering
\caption{Best-fit values to the trends shown in Figure \ref{fig:trends}.}
\label{table:line_fits}
\begin{tabular}{ccccc}
$z$ & Galaxies fit & s & b \\ \hline \hline \\[-5pt]
$\log{M}$ & \multicolumn{3}{c}{$(\log{r_{\rm{mass}} / r_{\rm{light}}} = s (\log{M/M_\odot} - 10) + b)$} \\ \hline \hline
% MASS
$1.0 \le z < 1.5$ & blue & $-0.143^{+0.017} _{-0.018}$ & $-0.097^{+0.008} _{-0.007}$ \\
$1.0 \le z < 1.5$ & red & $-0.042^{+0.021} _{-0.037}$ & $-0.084^{+0.026} _{-0.018}$ \\
$1.5 \le z < 2.0$ & blue & $-0.073^{+0.014} _{-0.015}$ & $-0.030^{+0.006} _{-0.006}$ \\
$1.5 \le z < 2.0$ & red & $-0.316^{+0.092} _{-0.069}$ & $0.088^{+0.035} _{-0.047}$ \\
$2.0 \le z < 2.5$ & blue & $-0.083^{+0.021} _{-0.022}$ & $0.004^{+0.006} _{-0.005}$ \\
$2.0 \le z < 2.5$ & red & $-0.145^{+0.163} _{-0.105}$ & $0.101^{+0.096} _{-0.149}$ \\[5pt]
% RE
$\log{r_{\rm{e,light}}}$ & \multicolumn{3}{c}{$(\log{r_{\rm{mass}} / r_{\rm{light}}} = s (\log{r_{\rm{e,light}}/{\rm{kpc}}} - 1) + b)$} \\ \hline \hline 
$1.0 \le z < 1.5$ & blue & $-0.325^{+0.073} _{-0.059}$ & $-0.233^{+0.026} _{-0.021}$ \\
$1.0 \le z < 1.5$ & red & $-0.103^{+0.027} _{-0.038}$ & $-0.180^{+0.024} _{-0.039}$ \\
$1.5 \le z < 2.0$ & blue & $-0.100^{+0.027} _{-0.029}$ & $-0.081^{+0.016} _{-0.020}$ \\
$1.5 \le z < 2.0$ & red & $-0.417^{+0.164} _{-0.117}$ & $-0.461^{+0.146} _{-0.096}$ \\
$2.0 \le z < 2.5$ & blue & $-0.034^{+0.029} _{-0.030}$ & $-0.029^{+0.016} _{-0.016}$ \\
$2.0 \le z < 2.5$ & red & $-0.187^{+0.124} _{-0.130}$ & $-0.186^{+0.125} _{-0.121}$ \\[5pt]
% N
n & \multicolumn{3}{c}{$(\log{r_{\rm{mass}} / r_{\rm{light}}} = s (\log n - 1) + b)$}\\ \hline \hline 
$1.0 \le z < 1.5$ & blue & $-0.198^{+0.054} _{-0.046}$ & $-0.110^{+0.009} _{-0.008}$ \\
$1.0 \le z < 1.5$ & red & $-0.045^{+0.051} _{-0.093}$ & $-0.092^{+0.052} _{-0.030}$ \\
$1.5 \le z < 2.0$ & blue & $0.014^{+0.023} _{-0.025}$ & $-0.032^{+0.006} _{-0.007}$ \\
$1.5 \le z < 2.0$ & red & $-0.185^{+0.187} _{-0.163}$ & $-0.039^{+0.080} _{-0.076}$ \\
$2.0 \le z < 2.5$ & blue & $-0.001^{+0.030} _{-0.033}$ & $-0.009^{+0.006} _{-0.007}$ \\
$2.0 \le z < 2.5$ & red & $0.016^{+0.151} _{-0.185}$ & $-0.038^{+0.107} _{-0.083}$ \\[5pt]
% U-V
$U-V$ & \multicolumn{3}{c}{$(\log{r_{\rm{mass}} / r_{\rm{light}}} = s ([U-V] - 1) + b)$} \\ \hline \hline
$1.0 \le z < 1.5$ & blue & $-0.170^{+0.016} _{-0.016}$ & $-0.106^{+0.011} _{-0.008}$ \\
$1.0 \le z < 1.5$ & red & $0.006^{+0.045} _{-0.113}$ & $-0.123^{+0.094} _{-0.038}$ \\
$1.5 \le z < 2.0$ & blue & $-0.103^{+0.014} _{-0.013}$ & $-0.052^{+0.007} _{-0.007}$ \\
$1.5 \le z < 2.0$ & red & $-0.520^{+0.233} _{-0.162}$ & $0.258^{+0.112} _{-0.151}$ \\
$2.0 \le z < 2.5$ & blue & $-0.091^{+0.021} _{-0.028}$ & $-0.031^{+0.010} _{-0.011}$ \\
$2.0 \le z < 2.5$ & red & $-0.376^{+0.348} _{-0.174}$ & $0.236^{+0.125} _{-0.223}$ \\[5pt]
% SIGMA
$\log{\Sigma_{\rm{eff}}}$ & \multicolumn{3}{c}{$(\log{r_{\rm{mass}} / r_{\rm{light}}} = s (\log{\Sigma_{\rm{eff}}} - 9) + b)$} \\ \hline \hline 
$1.0 \le z < 1.5$ & blue & $-3.452^{+0.279} _{-0.233}$ & $-27.968^{+2.253} _{-1.881}$ \\
$1.0 \le z < 1.5$ & red & $-2.504^{+0.579} _{-0.484}$ & $-20.192^{+4.632} _{-3.883}$ \\
$1.5 \le z < 2.0$ & blue & $-2.264^{+0.194} _{-0.196}$ & $-18.305^{+1.576} _{-1.587}$ \\
$1.5 \le z < 2.0$ & red & $-5.647^{+0.894} _{-0.737}$ & $-45.349^{+7.172} _{-5.925}$ \\
$2.0 \le z < 2.5$ & blue & $-2.022^{+0.455} _{-0.620}$ & $-16.303^{+3.668} _{-5.004}$ \\
$2.0 \le z < 2.5$ & red & $-6.623^{+0.942} _{-0.721}$ & $-53.028^{+7.505} _{-5.765}$ \\
\end{tabular}
\end{table}

While ${r_{\rm{mass}} / r_{\rm{light}}}$ does not appear to depend strongly on \sersic index, we do find that the color gradient strength correlates with stellar mass, half-light radius, $U-V$ color, and $\Sigma_{\rm{eff}}$ such that larger, more massive, and redder galaxies have more steeply negative color gradients. These trends are apparent both in the binned data and the values of the best-fit lines. We discuss the possible interpretations of these trends, as well as how they compare to previous studies, in Section \ref{sec:discussion}.

In Figure \ref{fig:ratioZtrend}, we show the median of ${r_{\rm{mass}} / r_{\rm{light}}}$ as a function of redshift for all galaxies with $M_* > 10^{10.08}M_\odot$. We see that both star-forming and quiescent galaxies in all redshift ranges typically have negative color gradients, where the outskirts of the galaxies are bluer--- and thus likely younger, more metal-poor, or less dusty--- than the centers of the galaxies. This is consistent with the histograms shown in the rightmost column of Figure \ref{fig:trends}, which show that from $1.0 < z < 2.0$, ${r_{\rm{mass}} / r_{\rm{light}}}$ is clearly not centered at one. This also agrees with many previous studies \citep[][]{tortora10,wuyts10,guo11,szomoru13,chan16,mosleh17}. Furthermore, we see evidence that color gradients are, on average, nearly flat at $z\gtrsim2$, then decrease steadily towards lower redshifts. Interestingly, quiescent and star-forming galaxies show no significant differences in their color gradient strength evolution. We explore the physical interpretation of this redshift evolution in Section \ref{sec:discussion}. 

We note that the trend shown in Figure \ref{fig:ratioZtrend} does not appear to be driven by differences in the average stellar masses of galaxies in each redshift bin; we see the same trend if we break the sample up into two mass bins of $10.0 \le \log{M/M_\odot} \le 10.5$ and $10.5 \le \log{M/M_\odot} \le 11.0$. This trend also does not appear to be driven by small galaxies, whose half-mass radii may be more difficult to recover: the redshift evolution for galaxies with $r_{\rm{light}} > 2\ \rm{kpc}$ shows the same trend as Figure \ref{fig:ratioZtrend}. Finally, we note that we see very similar trends between ${r_{\rm{mass}} / r_{\rm{light}}}$ and redshift when using any of the three methods of calculating half-mass radii described in Section \ref{sec:methods}.

\subsection{The galaxy mass - half mass radius relation}
\label{sec:mrrelation}
Because ${r_{\rm{mass}} / r_{\rm{light}}}$ varies with both stellar mass (Figure \ref{fig:trends} and redshift (Figure \ref{fig:ratioZtrend}), the galaxy $M_* - r_{\rm{mass}}$ relation differs from the galaxy $M_* - r_{\rm{light}}$ relation. Here, we show the galaxy $M_* - r_{\rm{mass}}$ relation for star-forming and quiescent galaxies at $1.0 \le z \le 2.5$.

\begin{figure*}[ht]
    \centering
    \includegraphics[width=.75\textwidth]{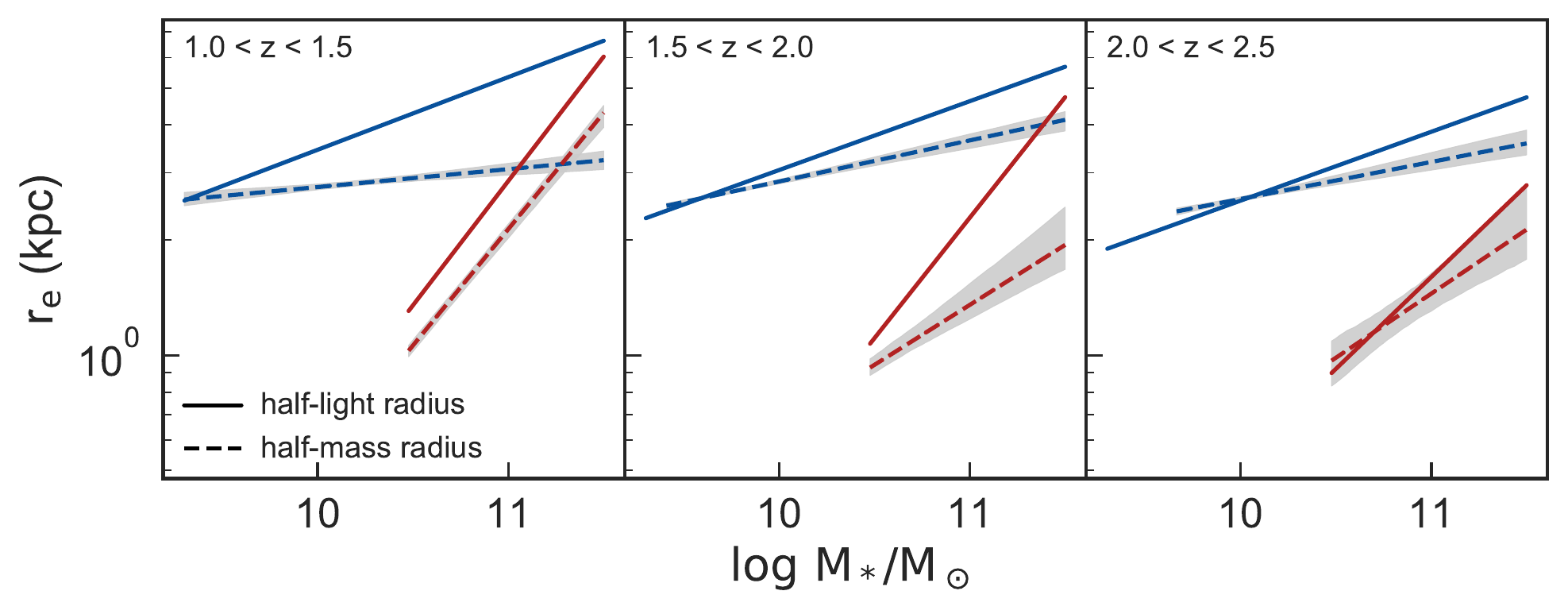}
    \caption{Mass-size relations for three redshift slices. The solid blue and red lines show the $M_* - r_{\rm{light}}$ relations from \citet{mowla18}, and the dashed lines show the $M_* - r_{\rm{mass}}$ relations from this analysis. Light grey regions show the 1$\sigma$ error region from 500 bootstrap simulations of the $M_* - r_{\rm{mass}}$ relation.}
    \label{fig:massSize}
\end{figure*}

Instead of directly fitting $r_{\rm{mass}}$ as a function of stellar mass, we `correct' the \citet{mowla18} mass-size relations from half-light radii to half-mass radii using our fits to ${r_{\rm{mass}} / r_{\rm{light}}}$ as a function of mass. We take this approach because here we are interested in how the relation {\it changes} when half-mass radii are considered instead of half-light radii. By using the \citet{mowla18} fits as the assumed true galaxy mass-light size relation, we minimize the effects that using a smaller sample size and a slightly different fitting technique would have on our conclusions. We note that we use the \citet{mowla18} mass-size relations over the \citet{vanderwel14} relations because the \citet{mowla18} fits use the same methods as \citet{vanderwel14}, but add a significant number of additional high-mass galaxies.\footnote{Both the half-mass radii presented here and the half-light radii used by \citet{mowla18} and \citet{vanderwel14} are semi-major axis measurements, not circularized radii.}

To obtain the $M_* - r_{\rm{mass}}$ relation, we simply multiply the \citet{mowla18} $r_{\rm{light}}-M_*$ relations by the ${r_{\rm{mass}} / r_{\rm{light}}}$ fits shown in Figure \ref{fig:trends} and Table \ref{table:line_fits}. The resulting  $r_{\rm{mass}}-M_*$ relations are shown in Figure \ref{fig:massSize}. Each panel shows a different redshift range. Solid lines represent the \citet{mowla18} $r_{\rm{light}}-M_*$ relations, while the dashed lines represent our $r_{\rm{mass}}-M_*$ relations. We only calculate the $r_{\rm{mass}}-M_*$ relations for masses where both our sample and the \citet{mowla18} sample are complete. The light grey region in Figure \ref{fig:massSize} shows the middle 68\% of 500 bootstrap realizations of the $r_{\rm{mass}}-M_*$ relation. 
Because galaxies tend to have smaller half-mass radii than half-light radii, the normalization of the $r_{\rm{mass}}-M_*$ relation is smaller than the normalization of the $r_{\rm{light}}-M_*$ relation for both star-forming and quiescent galaxies at all redshifts. The slope of both the star-forming and quiescent relations also clearly flattens at $1.0 \le z < 2.0$. In the $2.0 \le z \le 2.5$ bin, the slope of the star-forming relation also flattens; however, large uncertainties in the ${r_{\rm{mass}} / r_{\rm{light}}}$ fit for quiescent galaxies makes it difficult to draw conclusions for this population without a larger sample of galaxies. 
The slope of the $M_* - r_{\rm{mass}}$ relations are still approximately constant with redshift.

\begin{figure}[ht]
    \centering
    \includegraphics[width=.49\textwidth]{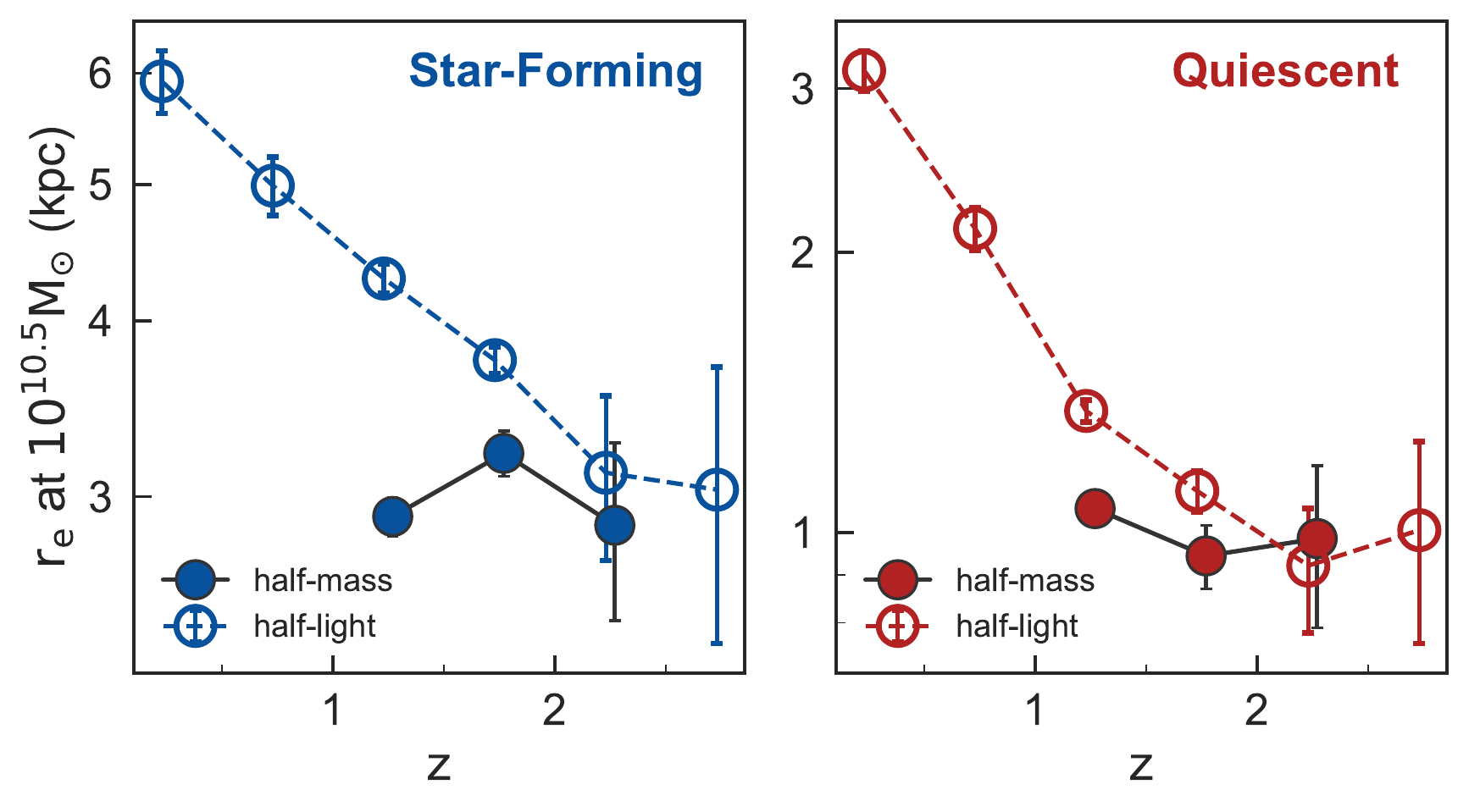}
    \caption{Half-light radii (open circles) and half-mass radii (filled circles) at $M_* = 10^{10.5}M_\odot$ for star-forming and quiescent galaxies as a function of redshift. Half-light radii are taken from the \citet{mowla18} fits, and half-mass radii are calculated using the fits in Table \ref{table:line_fits}. Over the $1.0 \le z \le 2.5$ range, the half-mass radii of both star-forming and quiescent galaxies evolve significantly less than their half-light radii.}
    \label{fig:rAtFixedM}
\end{figure}

In Figure \ref{fig:rAtFixedM}, we view the mass-size relation in a slightly different way: we use the best-fit mass-size relations to plot the radius at a fixed mass, $M_* = 10^{10.5} M_\odot$, as a function of redshift. The open circles show the values for the \citet{mowla18} fits to the $r_{\rm{light}}-M_*$ relation at $0.1 \le z \le 3.0$. The filled circles show the corresponding values of the $r_{\rm{mass}}-M_*$ relation fits (calculated from Table \ref{table:line_fits}, also shown in Figure \ref{fig:massSize}). Over the $1.0 \le z \le 2.5$ range studied in this paper, galaxy half-mass radii clearly evolve less than their half-light radii. While the half-light radii of star-forming galaxies at $M_* = 10^{10.5} M_\odot$ increase by  $1.17 ^{+0.41} _{-0.41}$~kpc ($37 ^{+19} _{-15}$\%) between $z=1.25$ and $z=2.25$, their half-mass radii only grow by $0.04 ^{+0.41} _{-0.42}$~kpc ($1 ^{+16} _{-13}$\%). Similarly, while the half-light radii of $M_* = 10^{10.5}M_\odot$ quiescent galaxies increase by $0.43 ^{+0.15} _{-0.13}$~kpc ($47 ^{+26} _{-19}$\%) over this redshift interval, their half-mass radii only grow by $0.08 ^{+0.21} _{-0.21}$~kpc ($8 ^{+29} _{-20}$\%). Over this redshift range, then, half-mass radii grow much less than half-light radii do. We note that the percentage growth in the half-mass and half-light radii of quiescent galaxies is consistent within the 1$\sigma$ error bars; a larger sample of high-redshift quiescent galaxies is required to tighten these error bars and further investigate galaxy size growth at high redshift.

%------------------------------------------------------------------------------------------------------------------------------------------------------------------------
\section{Discussion}
\label{sec:discussion}
\subsection{Sources of uncertainty}
Before providing a physical interpretation of the trends we see between color gradient strength and galaxy stellar mass, half-light radius, $U-V$ color, $\Sigma_{\rm{eff}}$, and redshift, we want to ensure that they do not arise from systematic uncertainties or biases in how well we are able to recover half-mass radii. 

The trends we see in color gradient strength (Figure~\ref{fig:trends}) are present regardless of which method we use to calculate half-mass radii. For both Method 1, Method 2, and Method 3, we find that ${r_{\rm{mass}} / r_{\rm{light}}}$ is nearly flat at $z\gtrsim2$ and decreases with decreasing redshift (Figure \ref{fig:ratioZtrend}); similarly, we see correlations between ${r_{\rm{mass}} / r_{\rm{light}}}$ and stellar mass, half-light radius, $U-V$ color, and $\Sigma_{\rm{eff}}$ for all three methods. The values of the best-fit trends for Methods 2 \& 3 are slightly different than those presented in Table \ref{table:line_fits} for Method 1, but generally agree within the error bars. While this agreement is expected because the three sets of half-mass radii are quite consistent (Figure \ref{fig:rvr}, Appendix~\ref{sec:methodtest}), it helps ensure that the trends we see are not simply a function of which method we chose to use.

All three of these methods, however, are based on the same multi-band imaging. One can imagine that it is simply easier to recover accurate mass profiles for galaxies which are physically larger on the sky--- because there are more pixels--- or more massive galaxies, which likely have higher signal-to-noise ratios. If this were the case, then color gradients in small or low-mass galaxies would be washed out and approach unity. These are the same trends that we see in the data. However, our methods do {\it not} appear to be significantly biased against recovering the half-mass radii of small or low-mass galaxies. In Appendix \ref{sec:recovery}, we test how well our primary method can recover the half-mass radii of modeled galaxies as a function of the properties of the modeled galaxy. We do not see any significant trends with half-light radius, stellar mass, or redshift. We do see that there is some dependence on color gradient, such that the half-mass radii of galaxies with negative color gradients tend to be overestimated more often than underestimated, and the half-mass radii of galaxies with positive color gradients tend to be underestimated more often than overestimated. In other words, the intrinsic color gradients are very slightly `washed out' for all galaxies. We note that this is a $\lesssim0.05$~dex effect; furthermore, correcting for this would only strengthen the trends that we see in ${r_{\rm{mass}} / r_{\rm{light}}}$.

Another potential bias in the imaging data is the PSF matching. The PSF of the F606W and F814W images is significantly smaller than the F125W, F140W, and F160W images. Therefore, if these images are not PSF-matched, galaxies will appear smaller in bluer bands, where the effect of the PSF is smaller. This will result in an artificial color gradient, such that the outskirts of a galaxy will seem redder than the center. We use PSF-matched images from the 3D-HST team \citep{skelton14} to mitigate this issue; however, errors in the PSF matching could similarly induce artificial color gradients. Figure 11 of \citet{skelton14} shows tests of the PSF matching, and they conclude that the PSF matching is accurate to 1\% for all bands and all fields. We have performed similar tests on the PSF-matched images using the same method we use to extract aperture photometry of galaxies, and find similar results. Therefore, it is unlikely that the color gradients we see in this paper are due to errors in the PSF matching.

When we fit the SEDs of each annulus in a galaxy, we fix the metallicity and leave only dust, age, and star formation timescale free. However, color gradients can also be caused by metallicity gradients--- by not allowing for metallicity gradients in our fits, we could be biasing the strength of the $M/L$ gradient. We chose to fix metallicity because allowing for a variable metallicity with only $5-8$ bands of resolved photometry results in poorly-constrained fits. The only parameter we use from the annuli SED fits is the stellar mass, which is generally considered the most well-constrained property from SED fits \citep[and is largely agnostic to the age-metallicity degeneracy, e.g.][]{bell01}.  
We are careful not to interpret the color gradients we see as gradients in any specific physical property, and focus on only the recovered half-mass radii. Significant additional work would be required to disentangle whether the gradients we see are due to age or metallicity. 

Another potential concern is the effect of dust. It is possible that we are systematically underestimating the masses of optically thick regions. Until we have spatially-resolved rest-frame mid- to far-IR data for a large number of galaxies, the best we can do is the integral constraint we use in this paper, where we ensure that the sum of the annuli SEDs matches the IRAC data points. We expect that the galaxies most likely to be affected by high dust obscuration are massive star-forming galaxies, whose centers are likely highly dust-obscured \citep[e.g.,][]{barro16}. By `missing' mass in the centers of these galaxies, we could be overestimating their half-mass radii and thus overestimating their ${r_{\rm{mass}} / r_{\rm{light}}}$. We note that this would result in an even  steeper trend in ${r_{\rm{mass}} / r_{\rm{light}}}$ as a function of mass than the one that we observe. 
We test the importance of this effect--- which, again, is likely strongest for dusty star-forming galaxies--- by removing all star-forming galaxies with $V-J > 1.2$ from our sample, re-fitting the trends shown in Figure \ref{fig:trends}, and re-calculating the effects on the mass-size relation shown in Figures \ref{fig:massSize} \& \ref{fig:rAtFixedM}. While the exact values change slightly (such that the implied growth in the half-mass radii of star-forming galaxies is 
$4_{-13}^{+17} \%$, instead of $\sim 1\%$), the overall trends we see are not significantly altered by excluding dusty star-forming galaxies from our sample.
Nonetheless, dust remains a significant uncertainty in this work, and our catalog of half-mass radii should be used with this consideration in mind.

Trends in ${r_{\rm{mass}} / r_{\rm{light}}}$ are also affected by uncertainties in $r_{\rm{light}}$ measurements. For this work, the most significant correction we make to $r_{\rm{light}}$ is to follow the \citet{vanderwel14} procedure to correct the measured F160W and F125W sizes to rest-frame 5,000\AA. This correction allows us to consistently compare half-light radii for galaxies at different redshifts. 
While it is important to apply this correction for consistency--- most other works in the literature apply this correction, including the \citet{vanderwel14} and \citet{mowla18} mass-size relation papers--- we note that it may affect trends in ${r_{\rm{mass}} / r_{\rm{light}}}$. For completeness, we have also studied the trends in ${r_{\rm{mass}} / r_{\rm{light}}}$ as a function of other galaxy properties if we take $r_{\rm{light}}$ to be the measured F160W size for all galaxies. All trends, at all redshifts, are still present.

We do not measure color gradients on scales smaller than the HWHM of the PSF. For small galaxies, then, we are only able to measure color gradients in the outer portions of the galaxy. 
Appendix \ref{sec:recovery} shows that we are still able to accurately recover the half-mass radii of small galaxies {\it if} their intrinsic $M/L$ profiles are power-law functions of radius. If the central regions of small galaxies deviate from this power-law $M/L$ profile, then the half-mass radii we recover for small galaxies may be biased. While we see evidence in this work \citep[and in previous works, e.g.][]{chan16} that a power law is a good model for the $M/L$ profiles of galaxies, higher-resolution imaging would be required to fully test this assumption for small galaxies.

Finally, our results--- and the results of all observational studies--- rely on the accuracy with which we can recover the intrinsic properties of galaxies from observations. \citet{price17} uses cosmological simulations of massive galaxies ($10^{10} - 10^{11.5}M_\odot$) at $1.7 \le z \le 2.0$ to assess how well observational techniques are able to recover stellar masses and half-mass radii. On average, they found that intrinsic half-mass radii were 0.1~dex smaller than observed half-mass radii as calculated with the  \citet{szomoru10,szomoru12,szomoru13} method. However, they found substantial scatter between the recovered and intrinsic properties; in particular, the viewing angle had strong effects on the scatter in the mass-size relation. The results presented here are likely subject to these effects.

In summary, we stress that the methods for calculating half-mass radii are complex, and in some respects lie towards the edge of what is possible with current data. Half-mass radii of individual galaxies should be treated with their error bars in mind. On the whole, however, we believe that the half-mass radii presented in this paper are not affected by strong systematic biases, and that the trends we see in ${r_{\rm{mass}} / r_{\rm{light}}}$ are a reflection of true trends in the data. 

\subsection{Comparison with previous studies}

The most basic trend we see is that color gradients in galaxies tend to be negative, such that the outskirts of galaxies are bluer than the central regions. We observe these generally negative color gradients for both star-forming and quiescent galaxies over the full $1.0\le z\le 2.5$ redshift range we study. This is in agreement with the consensus in the literature. Previous works explicitly measuring color gradients also find that they tend to be negative \citep[e.g.,][]{tortora10,wuyts10,guo11,szomoru13,chan16,mosleh17}. This effect is also seen by studies such as \citet{vanderwel14} that calculate galaxy sizes in multiple imaging bands: galaxies appear smaller if they are measured in longer-wavelength bands, implying negative color gradients. %It is heartening that our methods are able to reproduce this fairly basic observable.

The dependence of galaxy color gradients on other galaxy properties is more difficult to quantify; still, previous works have studied these correlations at both $z\sim0$ and $z\sim 1-2$. At $z\sim0$, \citet{tortora10} measured the color gradients of $\sim50,000$ galaxies in the SDSS. They find that color gradients are stronger for both higher-mass and larger galaxies, consistent with the trends we find at $1\le z\le2$. They additionally find that the strength of color gradients in quiescent galaxies seems to plateau at $\log M/M_\odot \ge 11$. Our data does not replicate this trend, perhaps due to the relatively low number of very high-mass quiescent galaxies in our sample or to genuine evolution of the high-mass quiescent population between $z\sim1$ and $z\sim0$. \citet{tortora10} attribute the trends they see mainly to metallicity gradients, while age gradients play a smaller but still significant role \citep[see also][]{greene12, greene15}. In this study, we are unable to separate the effects of age and metallicity. 

At $z>0$, previous studies have typically been hampered by small sample sizes, and have generally {\it not} found significant relationships between the strength of color gradients and other galaxy properties. For example, \citet{chan16} did not find significant correlations between ${r_{\rm{mass}} / r_{\rm{light}}}$ and stellar mass, size, or \sersic index in a sample of massive cluster galaxies at $z\sim1.4$.
\citet{szomoru13} found that ${r_{\rm{mass}} / r_{\rm{light}}}$ had no correlation with galaxy mass or surface density; they found very weak trends such that ${r_{\rm{mass}} / r_{\rm{light}}}$ was smaller for galaxies with higher \sersic indices, smaller sizes, and lower sSFRs. The discrepancy between the \citet{szomoru13} and \citet{chan16} results and the results presented in this paper are almost certainly due to large differences in sample size and mass range: the \citet{szomoru13} study only included 177 galaxies with $\log{M_*/M_\odot} > 10.7$, and the \citet{chan16} only studied 36 quiescent galaxies with $\log{M_*/M_\odot} > 10.2$; our analysis includes more than an order of magnitude more galaxies and spans two additional orders of magnitude in stellar mass. We note that if we consider only the quiescent galaxies in our sample with $\log{M_*/M_\odot} > 10.7$, we would see similarly weak or non-existant trends between color gradient strength and other galaxy properties. Our large sample size and long lever arm on stellar mass are thus crucial to see the clear dependence of ${r_{\rm{mass}} / r_{\rm{light}}}$ on both stellar mass, half-light radius, $U-V$ color, $\Sigma_{\rm{eff}}$, and redshift.

\citet{mosleh17} used CANDELS data to calculate the half-mass radii of $\sim2,000$ galaxies at $0.5 < z < 2.0$. Similar to our results, they find that the half-mass radii of star-forming galaxies evolve much more slowly than their half-light radii, and that color gradient strength depends on stellar mass. However, they find that the growth of quiescent galaxies is {\it not} significantly affected by color gradients. This discrepancy is likely due to differences in the methods used to calculate half-mass radii. \citet{mosleh17} find the best-fit single \sersic model for each band of imaging, then--- without correcting for residuals--- divide these fits into 1D spatial bins. They use SED fitting to find the mass in each bin, and fit the resulting mass profile with another \sersic model to extrapolate the mass profile to large radii. Using this method, the final half-mass radii depend strongly on the initial \sersic fits to each band of imaging data; errors on these fits do not appear to be included in the uncertainties on the final half-mass radii. Because this method is quite comparable to our Method 3 without the residual correction step, we do not implement it in this paper.

\subsection{Physical interpretation of the trends in color gradient strength}
In Section \ref{sec:trends}, we see that ${r_{\rm{mass}} / r_{\rm{light}}}$ decreases with increasing stellar mass, increasing half-light radius, increasing $U-V$ color, and increasing stellar mass surface density. Each of these trends is self-consistent in the framework of known galaxy correlations: more massive galaxies tend to be larger \citep[the mass-size relation, e.g.][]{shen03,vanderwel14,mowla18} and have redder colors \citep[e.g.][]{williams09}. $U-V$ color is also a proxy for sSFR--- galaxies redder in $U-V$ have lower sSFRs \citep[e.g.][]{williams09}. The main degeneracy when using $U-V$ as a sSFR proxy is between dusty star-forming galaxies and quiescent galaxies at the same $U-V$; our UVJ selection has effectively separated these populations. So, the trend in ${r_{\rm{mass}} / r_{\rm{light}}}$ as a function of $U-V$ tells us that, to first order, star-forming galaxies with lower sSFRs tend to have lower ${r_{\rm{mass}} / r_{\rm{light}}}$. Again, this fits into the framework of galaxy correlations: due to the shallow slope of the star-forming main sequence \citep[e.g.,][]{whitaker12} more massive galaxies tend to have lower sSFRs. 

While we do see that star-forming and quiescent galaxies occupy different regions of stellar mass, half-light radius, \sersic index, $U-V$, and $\Sigma_e$ space, the trends in ${r_{\rm{mass}} / r_{\rm{light}}}$ extend smoothly across the two populations. While the values for the best-fit trend lines (Table \ref{table:line_fits}) are somewhat different, the fits generally overlap within the error bars. This lack of bimodality in color gradient strength seems to suggest that color gradients may {\it gradually} become stronger as galaxies evolve from star-forming to quiescent.

Our findings are broadly consistent with the inside-out growth scenario. Before they quench, star-forming galaxies form new stars preferentially at large radii \citep{nelson16,hill17}, causing negative color gradients. 
Once they become quiescent, galaxies grow their sizes via minor mergers, which `puff up' the outskirts of the galaxy with bluer younger and/or lower-metallicity stars accreted in the mergers \citep[e.g.][]{bezanson09,naab09}. This would further steepen negative color gradients. This scenario could also explain why color gradients in quiescent galaxies are stronger at lower redshift: they have had more time to build up their bluer outer envelopes via mergers. Our basic findings are also consistent with other proposed quenching mechanisms that operate ``inside-out," such as quenching via wet compaction \citep[e.g.,][]{zolotov15}.

In a forthcoming paper, we will split our sample of galaxies by rest-frame SED shape \citep{kriek11} and study the strength of color gradients in different types of galaxies. By going beyond a simple `star-forming or quiescent' selection, we hope to test whether trends in color gradients change smoothly as galaxies evolve. We will also test whether recently-quenched galaxies have different half-mass radii or color gradients than older quiescent galaxies; this may provide clues to the physical mechanisms behind the processes that quench galaxies.

\subsection{Implications for the mass-size relation and its evolution with redshift}
The $r_{\rm{mass}}-M_*$ and $r_{\rm{light}}-M_*$ relations have several notable differences. Because galaxy color gradients tend to be negative, the intercept of the mass-size relation is smaller for half-mass radii than it is for half-light radii. This is true for both star-forming and quiescent galaxies at $1.0\le z\le2.5$. However, half-mass radii are not smaller than half-light radii by a consistent amount across all redshifts: high-redshift galaxies tend to have ${r_{\rm{mass}} / r_{\rm{light}}}\sim 1.0$, whereas lower-redshift galaxies tend to have ${r_{\rm{mass}} / r_{\rm{light}}}\sim 0.7$. This implies that there is {\it less evolution} in the $r_{\rm{mass}}-M_*$ relation than there is in the $r_{\rm{light}}-M_*$ relation. This trend can clearly be seen in Figure \ref{fig:rAtFixedM}, where we show that at $M_* = 10^{10.5}M_\odot$ and over the $1.0\le z\le 2.5$ redshift range, half-mass radii only increase by about $\sim$one-third the amount that half-light radii increase (however, this measurement has large error bars).

This slower size evolution sheds light on a longstanding question about the cause of the rapid size growth of quiescent galaxies at $z\gtrsim 1$. Both the growth of individual galaxies and growth of the population as a whole (i.e., progenitor bias) can contribute to observed size growth. However, several studies have shown that neither the growth of individual galaxies via minor mergers \citep[e.g.,][]{nipoti09,newman12,bedorf13} nor progenitor bias \citep[][]{belli15} alone is enough to account for the observed evolution in the half-light radii of galaxies between $z\sim2.5$ and $z\sim 1$. It appears that each method for size growth can only account for up to $\sim50\%$ of the observed increase in half-light radii \citep{newman12, belli15}. Here we show that the explanation for this remarkable size growth may simply be that the growth is not as rapid as was previously thought: the half-mass radii of quiescent galaxies at $M_* = 10^{10.5}M_\odot$ only increase by $8 ^{+29} _{-19}$\% between $z=1.25$ and $z=2.25$, less than the $47 ^{+25} _{-21}$\% increase in their half-light radii (Figure \ref{fig:rAtFixedM}).
It it therefore possible that minor mergers (or progenitor bias) {\it alone} is enough to account for the growth in the half-mass radii of quiescent galaxies. 

Because our current study focuses on the $1.0 < z < 2.5$ redshift range, we cannot address in detail how color gradients affect the size growth of quiescent galaxies below $z=1$ (Figure~\ref{fig:rAtFixedM}).  We can, however, use low-redshift color gradient studies to estimate the median half-mass radii of $z=0$ quiescent galaxies. We assume that $z=0$ quiescent galaxies have power-law $M/L$ profiles which decrease from $M/L_g\sim0.8$ to $M/L_g\sim0.3-0.4$ from 0--2.5$r_e$, as suggested by the $M/L_g$ profiles for E and S0 galaxies in Fig. 2 of \citet{garciabenito19}. We also assume that the light profiles of these galaxies are well-described by a $n\sim4.5$ \sersic profile (e.g., Mowla et al. 2018). Then, we
calculate the mass profiles and half-mass radii in a similar fashion to our Method 1. We estimate that ${r_{\rm{mass}} / r_{\rm{light}}} \sim 0.55$ for $z=0$ quiescent galaxies. This implies that color gradients may continue to evolve slightly below $z=1$. By applying this color gradient correction to the effective radii of low-redshift galaxies (Figure~\ref{fig:rAtFixedM}), we can calculate the effect on galaxy size evolution: between $z\sim2.5$ and $z\sim0$, the half-mass radii of galaxies may grow by only $\lesssim$~half of the amount that galaxy half-light radii grow.
However, we caution that this estimate is subject to large uncertainties--- the data sets, sample selection, and methods used to calculate color gradients at $z\sim0$ and $z>1$ are substantially different. Careful future analysis is required to fully understand the growth of half-mass radii over cosmic time. In a forthcoming paper, we will calculate the half-mass radii of $0.25 < z < 1.0$ galaxies in the CANDELS fields using the same methods presented in this paper. By using a uniform data set and methodology, we hope to understand in detail how quiescent galaxies grow over cosmic time.

The dependence of color gradient strength on redshift also has implications for star-forming galaxies: their half-mass radii at $M_* = 10^{10.5}M_\odot$ increase by only $1 ^{+16} _{-13}$\% between $z=1.25$ and $z=2.25$, significantly less than the $37 ^{+20} _{-15}$\% increase in their half-light radii. This may affect the nearly-linear scaling seen between the effective radii of galaxies and the virial radii of their host dark matter halos, and the resulting predictions for the redshift evolution of average disk effective radii \citep{kravtsov13, huang17, somerville17}. However, re-creating these abundance matching analyses using our catalog of half-mass radii is beyond the scope of this paper.

We note that--- like the $r_{\rm{light}}-M_*$ relations--- the star-forming and quiescent $r_{\rm{mass}}-M_*$ relations converge at the highest masses. Even when using half-mass radii, it is impossible to tell massive quiescent galaxies apart from massive star-forming galaxies by size alone. It is tempting to interpret the high-mass convergence of the two relations as part of an evolutionary sequence: perhaps one path to quiescence involves galaxies growing along the star-forming mass-size sequence until they reach a critical mass, after which they quench. This evokes the `parallel track' galaxy evolution model proposed in \citet{vandokkum15}. Our forthcoming paper will examine this in more depth by studying where galaxies with different rest-frame SED shapes lie in mass-size space, and how galaxies evolve through mass-size space to build up the star-forming and quiescent mass-size relations we see in Figure \ref{fig:massSize}.

%------------------------------------------------------------------------------------------------------------------------------------------------------------------------
\section{Conclusions}
In this paper, we present the largest study to date of galaxy color gradients and half-mass radii at $z>1$. We have tested three different methods for recovering galaxy half-mass radii from multi-band imaging. Two methods use a spatially-resolved SED modeling technique inspired by  \citet{wuyts12} to measure observed $M/L$ gradients. We then account for the effects of the PSF in two separate different ways: the first method uses a simple forward modeling technique that assumes the instrinsic $M/L$ gradient is a power-law function of radius, and the second technique uses \texttt{GALFIT} to fit a mass map of the galaxy. Our third and final method replicates the analysis of \citet{szomoru10,szomoru12,szomoru13} by using an intrinsic rest-frame color profile to infer a mass profile and half-mass radius. We find that all three methods produce remarkably consistent half-mass radii, with a scatter of 0.22~dex between the first two methods, and a scatter of $\sim0.3$~dex between those two methods and the \citet{szomoru10,szomoru12,szomoru13} method. The full catalog of galaxy half-mass radii calculated using each of these three techniques is released along with this paper (Table \ref{tab:results}).

With galaxy half-mass radii in hand, we study the strength of galaxy color gradients--- quantified by the ratio of half-mass to half-light radius ${r_{\rm{mass}}/r_{\rm{light}}}$--- as a function of various galaxy properties. We find that in general, both star-forming and quiescent galaxies have negative color gradients, such that they are redder in the centers and bluer on the outskirts. This result agrees with previous studies at both $z\sim0$ and $z>1$. The strength of these color gradients is, however, found to be a function of other galaxy properties. These trends have not been seen before in studies of color gradients at $z>1$; we are able to detect them due to our large sample size (more than an order of magnitude more galaxies than most previous studies) and large range of galaxy stellar masses. We  find that color gradients become more strongly negative as galaxies become more massive, larger, and redder.  Interestingly, these trends stretch across the star-forming and quiescent population smoothly without a clear bimodality. Furthermore, we measure significant evolution in the color gradient strength of both quiescent and star-forming galaxies as a function of redshift: color gradients are nearly flat at $z\gtrsim2$, then decrease steadily as redshift decreases. In total, ${r_{\rm{mass}}/r_{\rm{light}}}$ decreases by $\sim0.3$~dex between $z\sim2$ and $z\sim1$.

The observed trends appear consistent with the `inside-out growth' scenario \citep{bezanson09,naab09,nelson16}. Star-forming galaxies at these redshifts preferentially form stars in their outskirts; perhaps coupled with bulge growth, this would cause flat color gradients to become negative. Once galaxies quench and become quiescent, they grow their sizes via minor mergers, which introduce young and/or metal-poor stars to the galaxy outskirts; this also causes color gradients to decrease with time.

Finally, we used the dependence of ${r_{\rm{mass}}/r_{\rm{light}}}$ as a function of stellar mass and redshift to determine the effect of color gradients on the galaxy mass-size relation. Both the intercepts and the slopes of the quiescent and star-forming $r_{\rm{mass}}-M_*$ relations are smaller than those of the $r_{\rm{light}}-M_*$ relations. 
Moreover, we find that galaxy half-mass sizes grow {\it less rapidly} with redshift than half-light sizes: at $M_* = 10^{10.5}M_\odot$, the half-mass radii of star-forming and quiescent galaxies only grow by $0.04 ^{+0.42} _{-0.42}$ and $0.08 ^{+0.21} _{-0.20}$~kpc ($1 ^{+16} _{-13}$\% and $8 ^{+29} _{-19}$\%) between $z=2.25$ and $z=1.25$, whereas their half-light radii grow by $1.17 ^{+0.42} _{-0.42}$ and $0.43 ^{+0.15} _{-0.15}$~kpc ($37 ^{+20} _{-15}$\% and $47 ^{+25} _{-21}$\%). 

Further work remains to be done to understand galaxy half-mass radii and color gradients at $z>0$. In particular, in this study we are unable to determine the physical origin of galaxy color gradients, as we are not able to differentiate between the effects of age and metallicity. Dust is also a concern in this work: regions with very high attenuation may not be accounted for in any of the three methods we use, and our half-mass radii for very dusty galaxies should be treated with some caution. It is an open question whether the trends we see in color gradient strength as a function of redshift hold at lower or higher redshifts; examining color gradients over a wider redshift range could have further implications for the size growth of galaxies over cosmic time. Finally, the results presented here raise interesting questions as to how galaxies evolve and quench over the $1.0 < z < 2.5$ redshift range. 
In a future study, we will examine the color gradients and half-mass radii of galaxies as they transition from star-forming to quiescent.

\acknowledgements
We would like to thank Marijn Franx, Pieter van Dokkum, Arjen van der Wel, Lamiya Mowla, Rachel Bezanson, Tom Zick, and Jenny Greene for productive conversations.  We thank the anonymous referee for a helpful report. This work is based on observations taken by the CANDELS Multi-Cycle Treasury Program and the 3D-HST Treasury Program with the NASA/ESA {\it HST}, which is operated by the Association of Universities for Research in Astronomy, Inc., under NASA contract NAS5-26555. This work also uses data from the ZFOURGE survey; we thank the ZFOURGE team for making their data publicly available. This work is funded by grant AR-12847, provided by NASA though a grant from the Space Telescope Science Institute (STScI) and by NASA grant NNX14AR86G. This material is based upon work supported by the National Science Foundation Graduate Research Fellowship Program under grant No. DGE 1106400. K.A.S. also acknowledges support from
the University of California, Berkeley Chancellor's Fellowship.

\appendix 
\section{How well does Method 1 recover half-mass sizes?}
\label{sec:recovery}

In this Appendix, we assess whether Method 1 has significant systematic biases in recovering the half-mass radii of small, low-mass, or high-redshift galaxies.
We begin by taking the half-light radius, \sersic index, annuli boundaries, and $M/L$ gradient error bars for an observed galaxy in our sample. We use these quantities to model a hypothetical galaxy with a known half-mass radius: we assume that this galaxy has a `true' $M/L$ gradient parameterized as a power-law function of radius, and use the galaxy's light profile to calculate the resulting mass profile. We convolve this mass profile with the PSF and extract a convolved $M/L$ gradient. To ensure that we have realistic error bars, we assume that the error bars on this convolved $M/L$ gradient are the same as the error bars we measured for the real galaxy, with an unknown intrinsic $M/L$ profile. We randomly perturb the convolved $M/L$ gradient within these error bars to produce an `as-observed' $M/L$ gradient for this galaxy.

Then, we check how well Method 1 is able to recover the `true' $M/L$ gradient from the `as-observed' $M/L$ gradient. As described in Section \ref{sec:method1}, we perturb the galaxy's $n$ and $r_{\rm{e}}$, generate a set of possible convolved-space $M/L$ gradients, then use the  $\chi^2$ statistic to find the best-fit model. We then compare the recovered half-mass radius with the `true' half-mass radius that we chose for this galaxy.

\begin{figure}
    \centering
    \includegraphics[width=\textwidth]{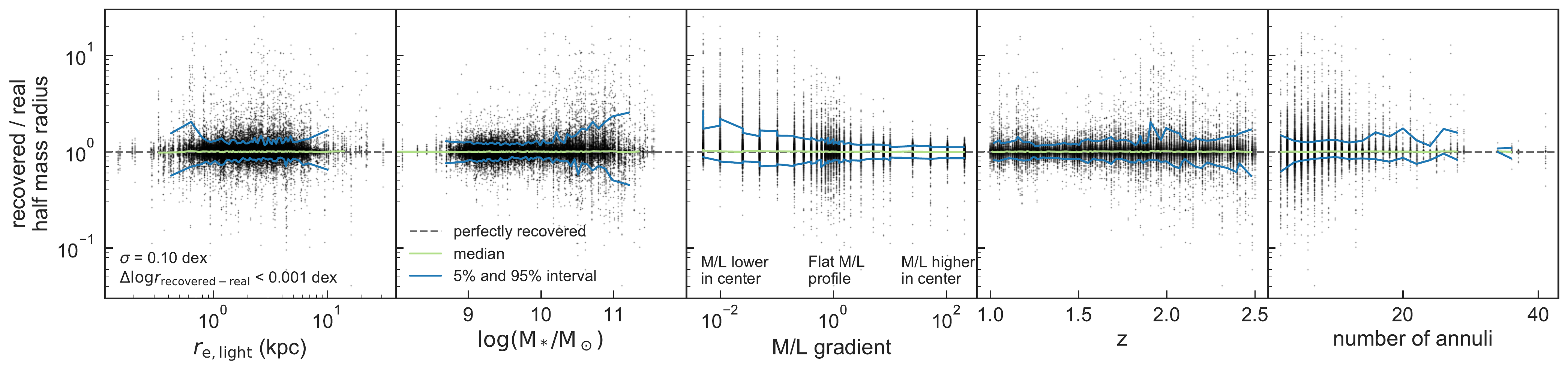}
    \caption{Ratio of recovered to real half-mass radius as a function of half-light radius, stellar mass, $M/L$ gradient, redshift, and number of annuli. The grey dashed line indicates where the recovered and real half-mass radii are equal. The green line shows the running median, and the blue lines show the 5\% and 95\% confidence intervals. In general, the scatter in the recovered half-mass radii are quite low, and there do not appear to be significant biases in how well Method 1 is able to recover the half-mass radii of small, low-mass, low-S/N, or high-redshift galaxies. Thus, the trends shown in Figures \ref{fig:trends} \& \ref{fig:ratioZtrend} are not explained by biases in how well our methods are able to recover half-mass radii. It does appear that the half-mass radii of galaxies with strongly decreasing $M/L$ gradients are slightly more difficult to recover than those with strongly increasing $M/L$ gradients.}
    \label{fig:recovery}
\end{figure}

We repeat this test using the $n$, $r_e$, annuli boundaries, and $M/L$ gradient error bars of every galaxy in our sample that lies in the COSMOS field. Additionally, for each galaxy we assume 27 different `true' $M/L$ gradient slopes covering the full range of $M/L$ slopes we include in Method 1. Figure \ref{fig:recovery} shows the ratio of the recovered half-mass radius to the `real' half-mass radius as a function of half-light radius, stellar mass, $M/L$ gradient slope, redshift, and number of annuli. The scatter between the recovered and true half-mass radii is $0.1$~dex, and the two measurements have an offset of $<0.001$~dex. 
There does not appear to be any significant trend in how well we are able to recover the half-mass radii with either half-light radius, stellar mass, redshift, or number of annuli. There is a slight trend with $M/L$ gradient such that the half-mass radii of galaxies with strongly decreasing $M/L$ gradients are slightly more difficult to recover than those with strongly increasing $M/L$ gradients; this $\sim0.05$~dex effect does not significantly affect our measurements.

\section{Do differences between Method 1 and Method 3 half-mass radii depend on galaxy properties?}
\label{sec:methodtest}
In Figure \ref{fig:rvr}, we show that the half-mass radii derived using three different methods generally agree. In this Appendix, we show in more detail that the differences between Method 1 and Method 3 half-mass radii do not depend on other galaxy properties studied in this paper. 

\begin{figure}
    \centering
    \includegraphics[width=.75\textwidth]{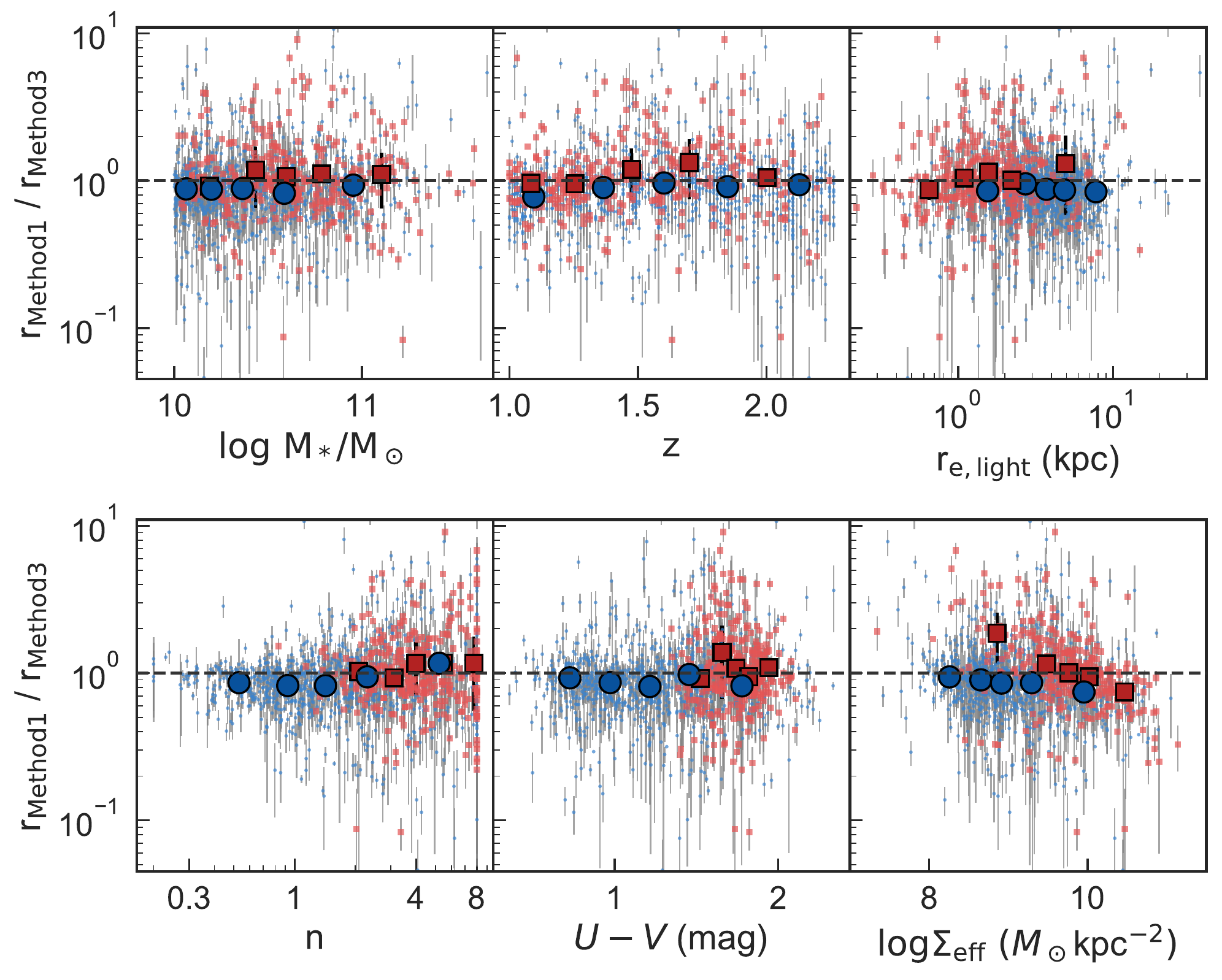}
    \caption{Ratio of Method 1 to Method 3 half-mass radii as a function of galaxy stellar mass, redshift, half-light radius, \sersic index, $U-V$ color, and $\Sigma_{\rm{eff}}$. We include all galaxies with stellar masses above $10^{10}M_\odot$ to ensure the sample is complete at all redshifts. Light blue points and light red squares represent individual galaxies. Large blue circles and red squares represent error-weighted means. The points are binned such that each mean point represents an equal number of individual data points.}
    \label{fig:app2}
\end{figure}

Figure \ref{fig:app2} shows the ratio of Method 1 to Method 3 half-mass radii as a function of other galaxy properties. While there is $\sim 0.3$~dex of scatter between the two methods of measuring half-mass radii, there are no significant trends with stellar mass, redshift, half-light radius, \sersic index, or $U-V$ color. There is a slight trend with $\Sigma_{\rm{eff}}$, such that Method 3 predicts higher half-mass radii than Method 1 for galaxies with higher $\Sigma_{\rm{eff}}$. If Method 3 half-mass radii were used in Figure \ref{fig:trends} instead of Method 1 half-mass radii, the trend with $\Sigma_{\rm{eff}}$ would be slightly less steep (but is still significant). All other trends would be essentially unchanged.

\bibliographystyle{aasjournal}
\bibliography{massBib}

\end{document}